\documentclass[review]{elsarticle}

\usepackage{hyperref}

\usepackage{amssymb}
\usepackage{amsfonts,amsmath,amsthm}
\usepackage{mathtools}
\usepackage{multirow}
\usepackage{fullpage} 
\usepackage{graphicx}
\usepackage{subcaption}
\usepackage{natbib}
\usepackage{xcolor}
\usepackage{multicol}
\usepackage{float}

\newcommand{\bpsi}{\boldsymbol{\psi}}
\newcommand{\dd}{\mathrm d}
\newcommand{\bX}{\boldsymbol{X}}
\newcommand{\by}{\mathbf y}

\newcommand{\bS}{\mathbf S}
\newcommand{\bs}{\boldsymbol{ s}}
\newcommand{\bz}{\boldsymbol{z}}
\newcommand{\bt}{\boldsymbol{ t}}
\newcommand{\bSigma}{\boldsymbol{\Sigma}}
\newcommand{\bomega}{\boldsymbol{\omega}}
\newcommand{\bell}{\boldsymbol{\ell}}

\newcommand{\Var}{\mathrm{Var}}
\renewcommand{\P}{\mathrm{P}}
\newcommand{\bspi}{\boldsymbol{\psi}}

\newcommand{\bR}{\mathbb{R}}
\newcommand{\sign}{\mathrm{sign}}

\newcommand{\rev}[1]{\textcolor{black}{#1}}

\newtheorem{theorem}{Theorem}

\biboptions{authoryear}

\begin{document}

\begin{frontmatter}

\title{\rev{A convolution type model for the intensity of spatial point processes applied to eye-movement data}}


\author[1,2]{Jean--François Coeurjolly \fnref{myfootnote}}
\ead{jean-francois.coeurjolly@univ-grenoble-alpes.fr}
\fntext[myfootnote]{Corresponding author}

\author[1,3]{Francisco Cuevas--Pacheco}
\author[1]{Marie--H\'el\`{e}ne Descary}

\address[1]{Department of Mathematics, Université du Québec à Montréal, Montréal, Canada}
\address[2]{University Grenoble Alpes, Grenoble-INP, LJK, 38000 Grenoble, France}
\address[3]{Departamento de Matem\'atica, Universidad T\'ecnica Federico Santa Mar\'ia, Avenida Espa\~na 1680, Valpara\'iso, Chile.}




\begin{abstract}
Estimating the first-order intensity function in point pattern analysis is an important problem, and it has been approached so far from different perspectives: parametrically, semiparametrically or nonparametrically. Our approach is close to a semiparametric one. Motivated by eye-movement data, we introduce a convolution type model where the log-intensity is modelled as the convolution of a function $\beta(\cdot)$, to be estimated, and a single spatial covariate (the image an individual is looking at for eye-movement data). Based on a Fourier series expansion, we show that the proposed model \rev{can be viewed as a} log-linear model with an infinite number of coefficients, which correspond to the spectral decomposition of $\beta(\cdot)$. After truncation, we estimate these coefficients through a penalized Poisson likelihood.
We illustrate the efficiency of the proposed methodology on simulated data and on eye-movement data.
\end{abstract}

\begin{keyword}
Point processes \sep Intensity function \sep Convolution \sep Fourier transform.
\end{keyword}

\end{frontmatter}


\section{Introduction}

Spatial point pattern data arise in many contexts where the interest lies in describing the distribution of an event in space. Some examples include the locations of trees in a forest, gold deposits mapped in a geological survey, stars in a cluster star \citep[see e.g.][]{moller2003statistical,illian2008statistical,baddeley2015spatial}. One of the main interests when analyzing spatial point pattern data is to estimate the intensity which characterizes the probability that a point (or an event) occurs in an infinitesimal ball around a given location. It is common to relate the distribution of points to one (or more) non-random spatial covariate $Z$ observed over the whole domain $W\subset \mathbb R^d$. This can be done semiparametrically \citep[][]{baddeley2015spatial} by assuming that $\rho(\bs)=f\{Z(\bs)\},\bs \in W,$ where $f:\bR \to \bR^+$ \rev{is estimated nonparametrically using a kernel method,} or using a parametric model, like the log-linear model 
\begin{equation} \label{eq:loglinear}
\log \rho (\bs) = \beta_0 + \beta Z(\bs), \quad \bs \in W,	
\end{equation}
where $\beta_0,\beta \in \bR$, which is certainly the most often used model. These models and in particular the latter one can easily be extended in a multivariate setting and have been applied in many various fields: e.g. to estimate intensity of species of trees \citep{waagepetersen2008estimating,choiruddin2020regularized}, of diseases locations in epidemiology surveillance \citep[see e.g.][]{gatrell1996spatial}, of locations of wildfire starts \citep[see e.g.][]{xu2011point}, etc.

The application we have in mind concerns eye-movement data which consist in locations of retina fixations recorded by an eye-tracker from individuals  looking at an image (or a video). It was observed \citep[see e.g.][]{Deubel,Wolfe-Horowitz,cerf,Judd} that fixations of the eyes are guided by local features of the image such as edges and colors, and by more global ones such as faces and objects. 
Parametric models have been investigated for example in \citet{barthelme2013modeling}. Sometimes, the temporal feature is also known and sequential spatio-temporal point processes models have been developed (see e.g. \citet{penttinen2016deducing} or \citet{ylitalo2016we}) to take this more complex situation into account.
 
From a biological point of view, when a human looks at an image, he explores locally a small part of the image. The resulting spatial average corresponds to the fixation. Then the eye of the subject jumps to another part of the image, explores it locally (in an imperceptible way for the human) and so on. In this paper, we assume we are only given a realization of a spatial point process (no temporal feature is observed) and
focus on a log-convolution model which relaxes the log-linear model \eqref{eq:loglinear} and which may take into account that spatial moving average characteristic. This new model is written as
\begin{equation}\label{eq:intensity}
\log \rho(\boldsymbol{s}) = \left( \beta*Z \right)(\boldsymbol{s}),     
\end{equation}
where the unknown parameter $\beta$ becomes a function in $L^{2}(\bR^{d})$, and $*$ denotes the convolution operator $(f*g)(\boldsymbol{s}) = \int_{\mathbb{R}^{d}}f(\boldsymbol{s} - \boldsymbol{\tau})g(\boldsymbol{\tau})\, \mathrm{d}\boldsymbol{\tau}$.
This model allows the intensity function evaluated at a location $\bs$ to depend not only on the single value $Z(\bs)$, like in the log-linear model, but also on all values of the covariate function $Z$. Indeed, $\log \rho(\boldsymbol{s})$ can be seen as an infinite weighted sum of the values of $Z$ over $W$, where the weights are defined through the function $\beta$. With respect to the previous application, $Z$ could correspond to the raw image (in gray level) or to the saliency map \citep[see e.g.][]{barthelme2013modeling} which is a prediction map of retina fixations built independently from the data.
 
Motivated by such an application, the aim of this paper is to consider the model~\eqref{eq:intensity} and to estimate nonparametrically the function $\beta(\cdot)$. Such a problem is close to a  deconvolution problem \citep[see][and the references therein]{starck2002deconvolution}. Therefore, we intend to provide a fast and efficient nonparametric estimator of the function $\beta(\cdot)$ by borrowing classical ideas from standard signal processing, image analysis and functional data. 

As detailed in Section~\ref{sec:background} our strategy is to take advantage of the Fourier basis which classically is able to handle efficiently convolutions. We decompose both $Z$ and $\beta$ in Fourier series and truncate these series. We show in Section~\ref{sec:background} that the resulting model will be close to a model of the form $\log \rho(\bs) \approx \boldsymbol{\psi}^\top \by (\bs)$, where $\boldsymbol \psi$ and $\by(\bs)$ are $p$-dimensional vectors obtained from Fourier decompositions of $\beta$ and $Z(\bs)$ respectively. This approximation is nothing else than a multivariate version of~\eqref{eq:loglinear} which is easily eastimated using standard methodology \citep[see e.g.][]{baddeley2015spatial,coeurjolly2014variational,guan2015quasi}. The length of the vector $\boldsymbol \psi$ to be estimated depends on the truncation of the Fourier series for $\beta$ and $Z$ and can get  large very quickly \rev{depending on the resolution of the image $Z$}. Hence, we borrow ideas from \citet{choiruddin2018convex} and implement a penalized composite likelihood. More specifically, we \rev{suggest to estimate the truncated spectrum of $\beta$ using}  adaptive Ridge and Lasso \rev{regularizations of the Poisson likelihood}. This is detailed in Section~\ref{sec:regularization}. Section~\ref{sec:simulation} presents a simulation study. In particular, we show that, for a large class of spatial point processes models, if the function $\beta$ has a sparse representation in the spectral domain, our adaptive Lasso estimator is able to estimate quickly and efficiently the Fourier coefficients of $\beta$ and the function $\beta$ itself by simply using inverse Fourier transform. 
\rev{Section~\ref{sec:data} presents an application to eye-movement data. We show that our approach has, on the one hand, excellent prediction performances, compared to other standard methods, and on the other hand, the interest to remain interpretable. Indeed, the function $\beta$ can be depicted and interpreted according to the image or conditions on the conducted experiment, etc.
Finally, even if we do not want to spotlight this result, we present, in \ref{sec:results}, infill asymptotic results for estimates derived from regularized versions of the Poisson likelihood. The infill characteristic is motivated by the application to image analysis: we assume observing more and more points in the same observation domain. This asymptotic result and its proof is quite similar to the ones obtained by~\citet{choiruddin2018convex} which considers increasing domain asymptotics.}

\section{\rev{Background, model approximation and estimation}} \label{sec:background}

\subsection{Background and notation}

Let $\bX$ be a point process defined on $\bR^{d}$, i.e., a random countable subset of $\bR^{d}$ observed in a bounded set $W \subset \bR^{d}$ which is typically a rectangular region and where usually $d=2,3$. Let $N(A)$ be the number of points of $\bX$ falling in $A \subset \bR^{d}$. We assume the intensity function associated with $\bX$ exists, that is the function $\rho: \bR^{d} \rightarrow \bR^{+}$ such that $\mathbb{E}\{N(A)\} = \int_{A}\rho(\bs) \, \mathrm{d} \boldsymbol{s}$ for $A \subset \mathbb{R}^{d}$ and $\rho(\boldsymbol{s}) > 0$ for all $\boldsymbol{s} \in W$ is well-defined.

Let $\mathbf{X}$ be a spatial point process on $\bR^d$. Let $W \subset \bR^d$ be a compact set of Lebesgue measure $|W|$ which will play the role of the observation domain. We view $\bX$ as a locally finite random subset of $\bR^d$, i.e. the random number of points of $\bX$ in $B$, $N(B)$, is almost surely finite whenever $B \subset \bR^d$ is a bounded region. A realization of $\bX$ in $W$ is thus a set $\mathbf{x}=\{x_1, x_2, \ldots, x_m\}$, where $x_i \in W$ and $m$ is the observed (finite) number of points in $W$.

Campbell theorem \citep[see e.g.][]{moller2003statistical} states that, assuming that $\bX$ has first-order (resp. second-order) intensity $\rho$ (resp. $\rho^{(2)}$) is equivalent to saying that for any function $k: \mathbb{R}^d \to [0,\infty)$ or $k: \mathbb{R}^d \times \mathbb{R}^d \to [0,\infty)$
\begin{align}
&\mathbb{E} \sum_{\bs \in \mathbf{X}} k(\bs)  ={\int_{\mathbb{R}^d} k(\bs) \rho (\bs)\mathrm{d}\bs} \label{eq:campbell} \\
&\mathbb{E} \sum_{\bs,\mathbf{t} \in \mathbf{X}}^{\neq} k(\bs,\mathbf t)=\int_{\mathbb{R}^d}{\int_{\mathbb{R}^d} k(\bs,\mathbf t) \rho^{(2)} (\bs,\mathbf t)\mathrm{d}\bs \mathrm{d}\mathbf t} \label{eq:campbell2}.
\end{align}
Therefore, for instance, we may interpret $\rho(u) \mathrm{d}u$ as the probability of occurrence of a point in an infinitesimally small ball with center $u$ and volume $\mathrm{d}u$. We end this section with the definition of the pair correlation function given by $g(\bs,\mathbf t) = \rho^{(2)}(\bs,\mathbf t)/\{\rho(\bs)\rho(\mathbf t)\}$ (with the convention $0/0=0$). This standard summary statistic measures the departure to the 'independence', as it is equal to 1 if $\bX$ corresponds to a Poisson point process.

\subsection{The log-convolution model and its approximation}

Let us first describe more rigorously the model~\eqref{eq:intensity}. We assume that the covariate $Z: \bR^{d} \rightarrow \bR$ is a function of $L^2(\bR^{d})$ the space of square integrable functions equipped with the inner product $\langle f,g \rangle = \int_{\bR^{d}}f(\boldsymbol{s}) g(\boldsymbol{s}) d\, \boldsymbol{s}$. We assume without loss of generality that the process $\bX$ is observed on the hypercube $W = [0 ,1]^d$, and that both $\beta$ and $Z$ are functions with support $W$. Therefore it is possible to expand $\beta$ and $Z$ using a Fourier basis of $L^2(W)$. Indeed, let $\phi_{\boldsymbol{\kappa}}(\boldsymbol{s}) = \exp(2 \pi \mathrm{i} \boldsymbol{\kappa} \cdot \boldsymbol{s})=\prod_{i = 1}^{d}\exp(2\pi \mathrm{i} \kappa_i s_i)$ be the Fourier bases, where $\mathrm{i}$ is the imaginary unit, $\boldsymbol{s}=\rev{(s_1,\ldots,s_d)^\top} \in W$, and $\boldsymbol{\kappa} = \rev{(\kappa_1, \ldots, \kappa_d)^\top}\in \mathbb{Z}^{d}$ is a vector of frequencies, then we can write

\begin{equation}\label{eq:fourier}
\beta(\boldsymbol{s}) = \sum_{\boldsymbol{\kappa} \in \mathbb{Z}^{d}} \beta_{\boldsymbol{\kappa}}\phi_{\boldsymbol{\kappa}}(\boldsymbol{s}), \qquad Z(\boldsymbol{s}) = \sum_{\boldsymbol{\kappa} \in \mathbb{Z}^{d}} Z_{\boldsymbol{\kappa}}\phi_{\boldsymbol{\kappa}}(\boldsymbol{s}),
\end{equation}

\noindent where $\beta_{\boldsymbol{\kappa}} = \langle \beta,\phi_{\boldsymbol{\kappa}} \rangle$ and $Z_{\boldsymbol{\kappa}} = \langle Z,\phi_{\boldsymbol{\kappa}} \rangle$ correspond to the $\boldsymbol{\kappa}$-Fourier coefficients of $\beta$ and $Z$ respectively. By replacing the functions $\beta$ and $Z$ in the log-convolution model \eqref{eq:intensity} by their expression given in \eqref{eq:fourier}, we have


\begin{equation}\label{eq:integration}
    \log \rho(\boldsymbol{s}) = \left( \beta*Z \right)(\boldsymbol{s}) = \sum_{\boldsymbol{\kappa} \in \mathbb{Z}^{d}} \sum_{\boldsymbol{l} \in \mathbb{Z}^{d}} {\beta}_{\boldsymbol{\kappa} }{Z}_{\boldsymbol{l}} \int_W \phi_{\boldsymbol{\kappa}}(\bs-\boldsymbol{\tau})\phi_{\boldsymbol{l}}(\boldsymbol{\tau})d\boldsymbol{\tau}= \sum_{\boldsymbol{\kappa} \in \mathbb{Z}^{d}} {\beta}_{\boldsymbol{\kappa} }{Z}_{\boldsymbol{\kappa}}\phi_{\boldsymbol{\kappa}}(\boldsymbol{s}),
\end{equation}
where the last equality is obtained using the fact that $\phi_{\boldsymbol{\kappa}}(\bs-\boldsymbol{\tau})=\phi_{\boldsymbol{\kappa}}(\bs)\phi_{\boldsymbol{\kappa}}(-\boldsymbol{\tau})=\phi_{\boldsymbol{\kappa}}(\bs)\overline{\phi_{\boldsymbol{\kappa}}(\boldsymbol{\tau})}$, with $\overline{z}$ denoting the complex conjugate of $z$. Hence, our model can be rewritten in terms of the spectrum of $\beta$ and $Z$. Note that since $\log \rho(\boldsymbol{s})$ is a real number, its imaginary part $\mathcal{I}[\log \rho(\boldsymbol{s})]$ must be zero. Moreover, since the Fourier coefficients of a real function satisfy the Hermitian symmetry property, then $\beta_{\boldsymbol{\kappa}} = \overline{\beta_{-\boldsymbol{\kappa}}}$ and $Z_{\boldsymbol{\kappa}} = \overline{Z_{-\boldsymbol{\kappa}}}$, for $\boldsymbol{\kappa}\in  \mathbb{Z}^{d}$. Defining $\mathbb{Z}^{d}_{\oplus}$ as the set of unique non-zero vectors $\boldsymbol{\kappa}\in \mathbb{Z}^{d}$ under the relation $\boldsymbol{\kappa} = -\boldsymbol{\kappa}$, and $\boldsymbol{\kappa}_0=\boldsymbol{0}$ as the d-dimensional zero vector (yielding to $\phi_{\boldsymbol{\kappa}_0}(\boldsymbol{s})=\exp(2 \pi \mathrm{i} \boldsymbol{0} \cdot \boldsymbol{s})=1$), we can rewrite (\ref{eq:integration}) as

\begin{eqnarray}
    \log \rho(\boldsymbol{s}) &=& \beta_{\boldsymbol{\kappa}_0}Z_{\boldsymbol{\kappa}_0}\phi_{\boldsymbol{\kappa}_0}(\boldsymbol{s})+ \sum_{\boldsymbol{\kappa} \in \mathbb{Z}^{d}_{\oplus}} \left\{ \beta_{\boldsymbol{\kappa}}Z_{\boldsymbol{\kappa}}\phi_{\boldsymbol{\kappa}}(\boldsymbol{s}) + \overline{\beta_{\boldsymbol{\kappa}}Z_{\boldsymbol{\kappa}}\phi_{\boldsymbol{\kappa}}(\boldsymbol{s})} \right\}\nonumber \\
 &=& \beta_{\boldsymbol{\kappa}_0}Z_{\boldsymbol{\kappa}_0}+ \sum_{\boldsymbol{\kappa} \in \mathbb{Z}^{d}_{\oplus}}  2\mathcal{R}[\beta_{\boldsymbol{\kappa}}Z_{\boldsymbol{\kappa}}\phi_{\boldsymbol{\kappa}}(\boldsymbol{s})]  \nonumber \\
 &=& \beta_{\boldsymbol{\kappa}_0}Z_{\boldsymbol{\kappa}_0}+ \sum_{\boldsymbol{\kappa} \in \mathbb{Z}^{d}_{\oplus}} \left\{2 \mathcal{R}[\beta_{\boldsymbol{\kappa}}]\mathcal{R}[Z_{\boldsymbol{\kappa}}\phi_{\boldsymbol{\kappa}}(\boldsymbol{s})] - 2\mathcal{I}[\beta_{\boldsymbol{\kappa}}]\mathcal{I}[Z_{\boldsymbol{\kappa}}\phi_{\boldsymbol{\kappa}}(\boldsymbol{s})] \right\},
 \label{eq:conjugate}
\end{eqnarray}

\noindent with $\mathcal{R}[z]$ being the real part of $z$, and where the identities $z+\bar{z}=2\mathcal{R}[z]$ and $\mathcal{R}[z_1z_2] = \mathcal{R}[z_1]\mathcal{R}[z_2] - \mathcal{I}[z_1]\mathcal{I}[z_2]$ have been used to obtain the second and last equality respectively. In the sequel we denote $\mathcal{R}[\beta_{\boldsymbol{\kappa}}]$ (respectively $\mathcal{I}[\beta_{\boldsymbol{\kappa}}]$) by $\beta_{\boldsymbol{\kappa}}^R$ (respectively $\beta_{\boldsymbol{\kappa}}^I$).

The log-convolution model (\ref{eq:conjugate}) cannot be used directly since as already pointed out, it depends on the covariate function $Z$ through its spectrum which is not observed in practice. We propose to  estimate it in  an efficient way by using the $d$-dimensional fast Fourier transform (FFT) of the sequence $\{Z(\boldsymbol{s}_{i})\}_{i=1}^{N}$, where $\{\boldsymbol{s}_{i}\}_{i=1}^{N}$ is a regular grid over $W$. The resulting approximation of the spectrum has a finite number of terms $\{{Z}_{\boldsymbol{\kappa}_i}\}_{i=0}^K$, where $\{\boldsymbol{\kappa}_i\}_{i = 0}^{K}$ is a partially ordered sequence of the obtained Fourier frequencies. This yields the following finite approximation of the model (\ref{eq:conjugate}):

\begin{equation}\label{eq:linear_model1}
\log \rho_K(\boldsymbol{s}) =\beta_{\boldsymbol{\kappa}_0}Z_{\boldsymbol{\kappa}_0}+  \sum_{i = 1}^{K} \left\{2\beta_{\boldsymbol{\kappa}_{i}}^{R}\mathcal{R}[Z_{\boldsymbol{\kappa}_{i}}\phi_{\boldsymbol{\kappa}_i}(\boldsymbol{s})] - 2\beta_{\boldsymbol{\kappa}_{i}}^{I}\mathcal{I}[Z_{\boldsymbol{\kappa}_{i}}\phi_{\boldsymbol{\kappa}_i}(\boldsymbol{s})] \right\}.\\
\end{equation}

\noindent Note that since $\phi_{\boldsymbol{\kappa}_0}(\boldsymbol{s})=1$, we have that
\begin{equation*} 
 \beta_{\boldsymbol{\kappa}_0} = \langle \beta,\phi_{\boldsymbol{\kappa}_0} \rangle = \int_{W} \beta(\boldsymbol{s}) \, \rm{d}\boldsymbol{s} \textrm{ and }  Z_{\boldsymbol{\kappa}_0} = \langle Z,\phi_{\boldsymbol{\kappa}_0} \rangle = \int_{W} Z(\boldsymbol{s}) \, \rm{d}\boldsymbol{s} ,
\end{equation*}
are two real quantities. Moreover, from expression \eqref{eq:linear_model1}, we note that the parameter $\beta_{\boldsymbol{\kappa}_{0}}$ is not identifiable if $Z_{\boldsymbol{\kappa}_0} = 0$. To avoid this problem, we suppose without loss of generality that $Z_{\boldsymbol{\kappa}_0}=\int_{W} Z(\boldsymbol{s}) \, \rm{d}\boldsymbol{s}=1$. 
Finally, we can write the finite approximation \eqref{eq:linear_model1} of our original log-convolution model \eqref{eq:intensity} as a log-linear model 
\begin{equation} \label{eq:linear_model3}
\log \rho_K(\boldsymbol{s}) = \beta_{\boldsymbol{\kappa}_{0}}+\boldsymbol{\beta}_K^\top\boldsymbol{Z}(\boldsymbol{s}),
\end{equation}
where $\beta_{\boldsymbol{\kappa}_{0}} \in \mathbb{R}$ and $\boldsymbol{\beta}_{K}=(\beta^{\mathcal{R}}_{\boldsymbol{\kappa}_{1}},\ldots,\beta^{\mathcal{R}}_{\boldsymbol{\kappa}_{K}},\beta^{\mathcal{I}}_{\boldsymbol{\kappa}_{1}},\ldots,\beta^{\mathcal{I}}_{\boldsymbol{\kappa}_{K}})^\top \in \mathbb{R}^{2K}$ are the parameters to be estimated and $\boldsymbol{Z}(\boldsymbol{s})$ is the vector of covariates defined as
\begin{equation*}
 \boldsymbol{Z}(\boldsymbol{s}) = \{2\mathcal{R}[Z_{\boldsymbol{\kappa}_{1}}\phi_{\boldsymbol{\kappa}_{1}}(\boldsymbol{s})], \ldots, 2\mathcal{R}[Z_{\boldsymbol{\kappa}_{K}}\phi_{\boldsymbol{\kappa}_{K}}(\boldsymbol{s})], -2\mathcal{I}[Z_{\boldsymbol{\kappa}_{1}}\phi_{\boldsymbol{\kappa}_{1}}(\boldsymbol{s})], \ldots, -2\mathcal{I}[Z_{\boldsymbol{\kappa}_{K}}\phi_{\boldsymbol{\kappa}_{K}}(\boldsymbol{s})] \}^\top.  
    \end{equation*}
In the following, we assume that the approximation is exact that is there exists $K<\infty$ (potentially very large) such that $\rho=\rho_K$. Given the form of~\eqref{eq:linear_model3}, it is therefore possible to estimate the spectrum of the function $\beta$ using existing estimation methods for log-linear models, given that the covariate function $Z$ is such that there is at least one $\boldsymbol{\kappa}_i \in \mathbb{Z}^{d}, i=1,\ldots,K$ such that $Z_{\boldsymbol{\kappa}_i}$ is non-zero. Finally, once the spectrum of $\beta$ is estimated, we can get back to the function $\beta$ using the inverse fast Fourier transform. \rev{Therefore, the estimation problem results in estimating $\beta_{\boldsymbol{\kappa}_{0}}$ and $\boldsymbol{\beta}_K$ quickly and efficiently. Given the fact that $\boldsymbol{\beta}_K$ can be a high-dimensional vector, we have to adapt the estimation procedure to this large dimension context. This is investigated in the next paragraph.}


\subsection{Regularized estimation procedure}\label{sec:regularization}

To avoid any confusion with previous paragraphs, we slightly change our notation in this section. The model \eqref{eq:linear_model3} \rev{is a particular case of~\eqref{eq:model}}.
We assume to observe a sequence of  spatial point processes $\bX_n$ with intensity $\rho_n$ parameterized by $\bpsi \in \boldsymbol{\Psi} \subseteq \mathbb{R}^p$ ($p\ge 1$) as
\begin{equation}
    \label{eq:model}
    \log \rho_n(\bs)  = \theta_n + \bpsi^\top \by(\bs), \quad \bs \in W.
\end{equation}
Here the parameter $\theta_n$ should be regarded as a nuisance parameter and is such that $\theta_n\to \infty$ as $n \to \infty$ while $\by(\bs) = (y_1(\bs),\dots, y_p(\bs))^\top$ represents the vector of $p$ spatial covariates available at any location $\bs\in W$. As a matter of fact, $W$ is assumed to be fixed and the mean number of points in $W$ which is proportional to $\theta_n$ grows with $n$. This setting, called infill asymptotics in the spatial statistics literature, is natural in the context of the present paper (the image is fixed). 

The inference is done by maximizing $Q_n(\cdot)$, an adaptive Lasso regularized version of the Poisson likelihood \citep{choiruddin2020regularized}, namely
\begin{equation}
    \label{eq:Qn}
    Q_n(\bpsi) = \theta_n^{-1} \ell_n(\bpsi) - \sum_{j=1}^p \lambda_{n,j} |\psi_j|,
\end{equation}
where
\begin{equation}
    \label{eq:elln}
    \ell_n(\bpsi) = \sum_{\bs\in \bX_n} \log \rho_n(\bs)  -\int_W \rho_n(\bs) \dd \bs
\end{equation}
and where $\lambda_{n,j}\ge 0$ are regularization parameters. \rev{\ref{sec:results} provides conditions on the model (essentially on covariates and dependence characteristics of $\bX_n$) and on the sequence of regularization parameters such that the estimate $\hat{\boldsymbol{\psi}} = \mathrm{argmin}_{\boldsymbol{\psi}} Q_n(\boldsymbol{\psi})$ is consistent, sparse and satisfies a central limit theorem. In the simulation and data analysis, we have also considered the ridge penalty which consists in replacing $|\psi_j|$ by $\psi_j^2/2$ in~\eqref{eq:Qn}. Note that the adaptive ridge estimator is consistent but does not satisfy any sparse property.}

\rev{It is well-known that the quality of the estimation obtained using adaptive Lasso or adaptive Ridge relies on the sequence $\{\lambda_{n,j}\}$ \citep{fan2010selective}. We follow \citet{zou2006adaptive} and \citet{choiruddin2018convex} and set $\lambda_{n,j} = \lambda/|\hat{\psi}_{j}^R|$ where $\hat{\psi}_{j}^R$ is a preliminary ridge estimation of $\psi_j$. This transforms the original problem of finding a sequence $\{\lambda_{n,j}\}_{j = 1}^{p}$ into the one of choosing the parameter $\lambda$. This has been done by minimizing \rev{a version of the composite BIC criterion \citep{choiruddin2018convex,choiruddin2021information,ba2020high} adapted to large dimensional problems
\begin{equation*}
    \mathrm{CBIC}(\lambda) = -2 \ell\{ \hat{\bpsi}(\lambda) \} + s(\lambda)\log n(W),
\end{equation*}
where $n(W)$ is the observed number of data points and where} $\hat{\bpsi}(\lambda)$ is the estimate of $\{\psi_j\}_{j = 1}^{p}$ for a given $\lambda$ and $s(\lambda)$ is the total number of non-zero parameters. The minimization of $\mathrm{CBIC}$ is performed using a grid search method on the interval $[\lambda_{\min},\lambda_{\max}]$ where $\lambda_{\min}$ and $\lambda_{\max}$ are obtained following \citet{friedman2010regularization}.}

\section{Simulation study}\label{sec:simulation}

In this section, we investigate the finite-sample properties of the adaptive Lasso estimator introduced in Section \ref{sec:results} through a simulation study. We  also compare this estimator with an adaptive Ridge estimator and the standard Poisson likelihood estimator.
 
\begin{figure}[H] 
\includegraphics[scale = 0.25]{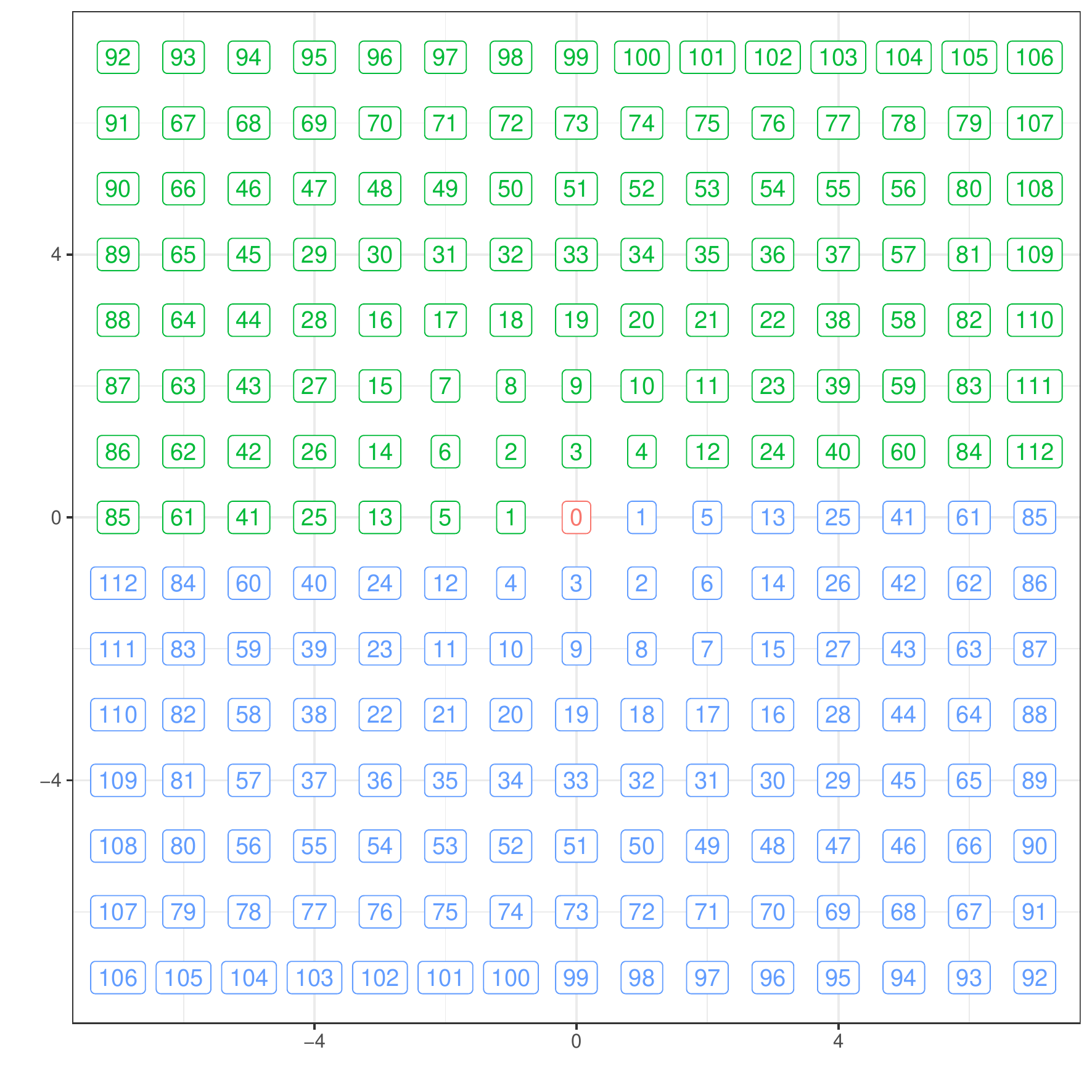} \includegraphics[scale=0.5]{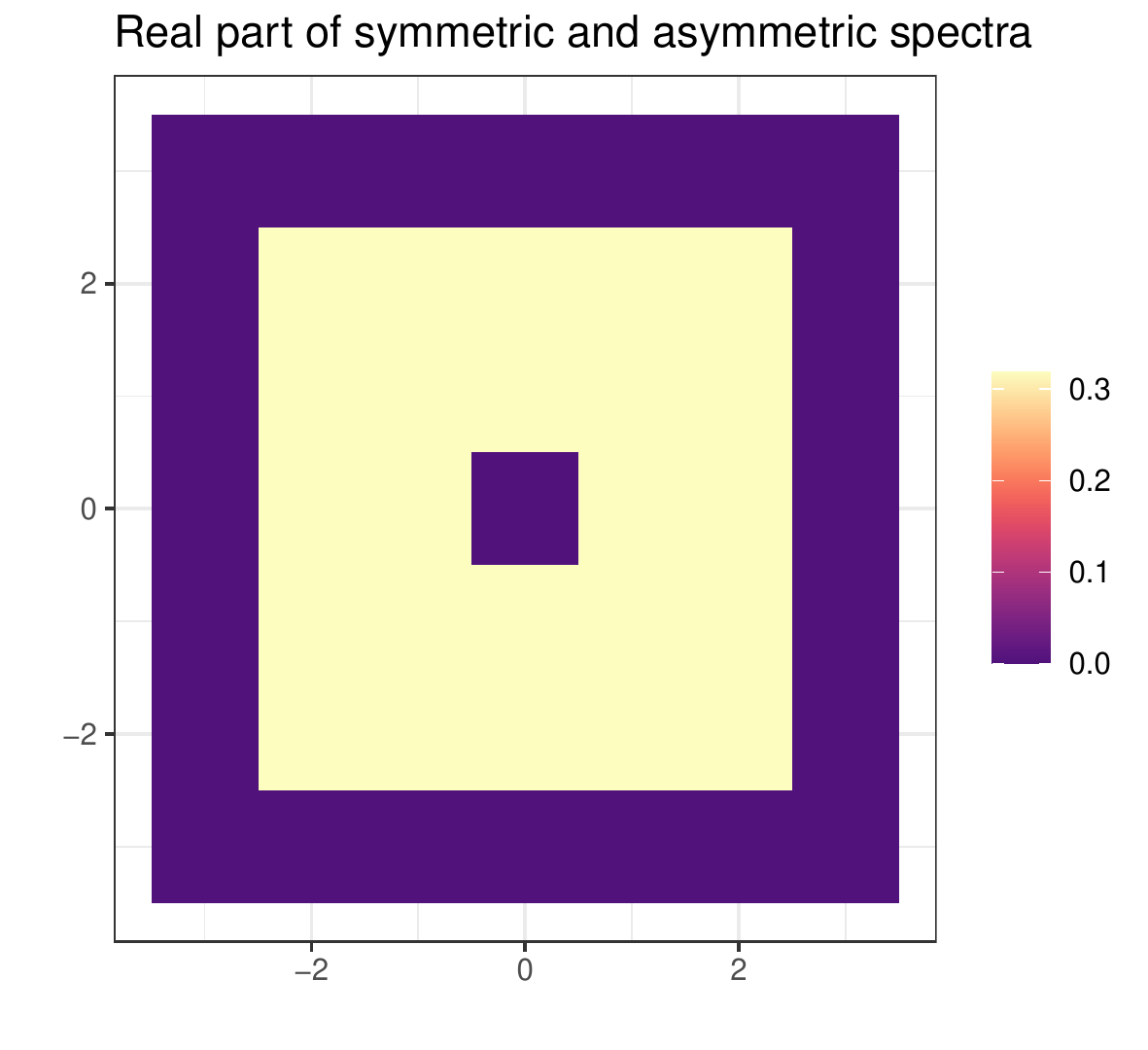}\includegraphics[scale=0.5]{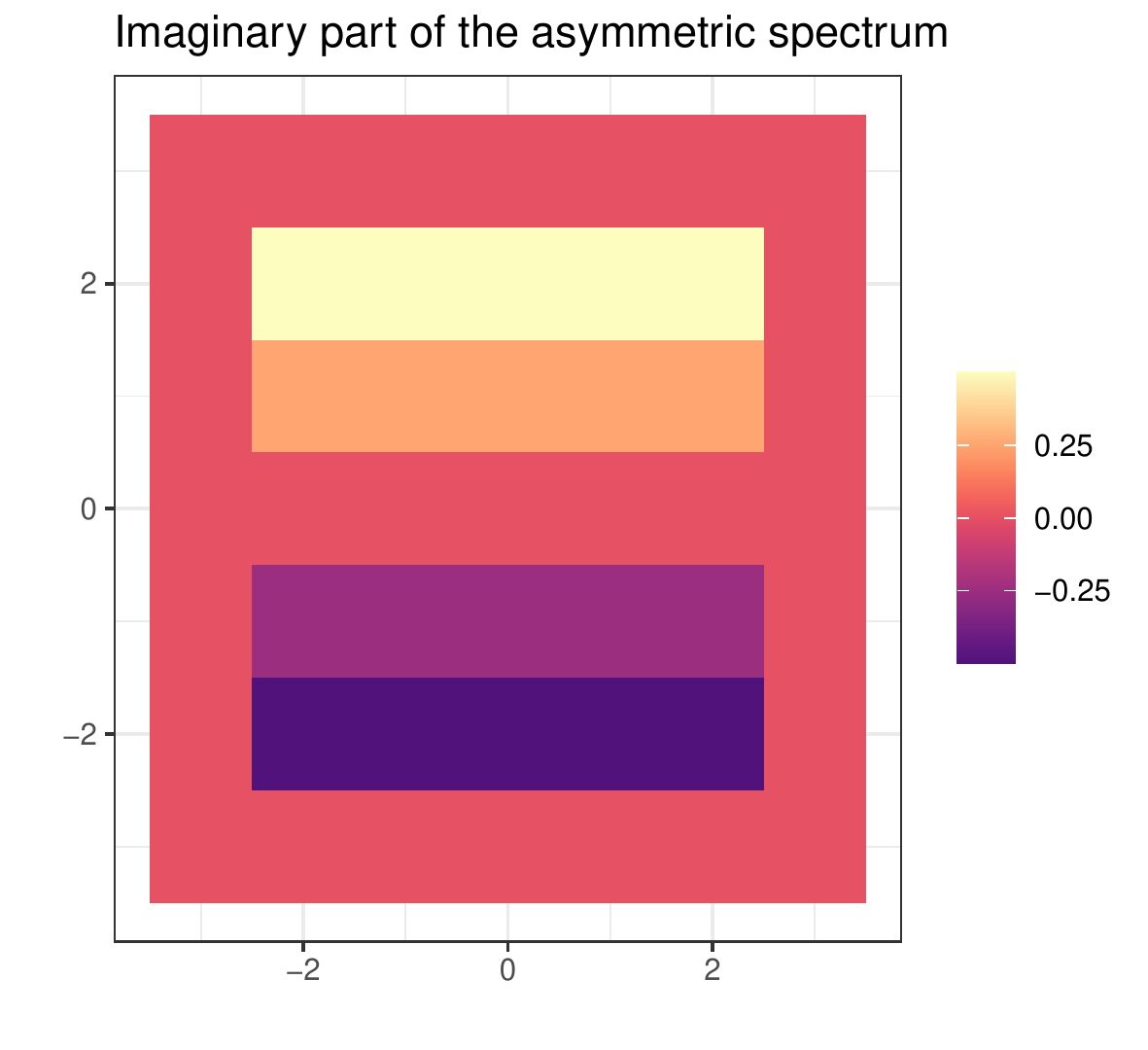}
\caption{(Left) Partial order of the sequence of frequencies. The green boxes are the used frequencies and the blue boxes are the conjugate ones. The zero frequency $\boldsymbol{\kappa}_0=(0,0)$ is always included. (Middle and right) Real and imaginary part of the spectrum of $\beta$ described in (a) and (b) of Section \ref{sec:simulation}.} \label{fig:spectrum} 
\end{figure}

\begin{figure}[H] 
\centering
\includegraphics[scale=.8]{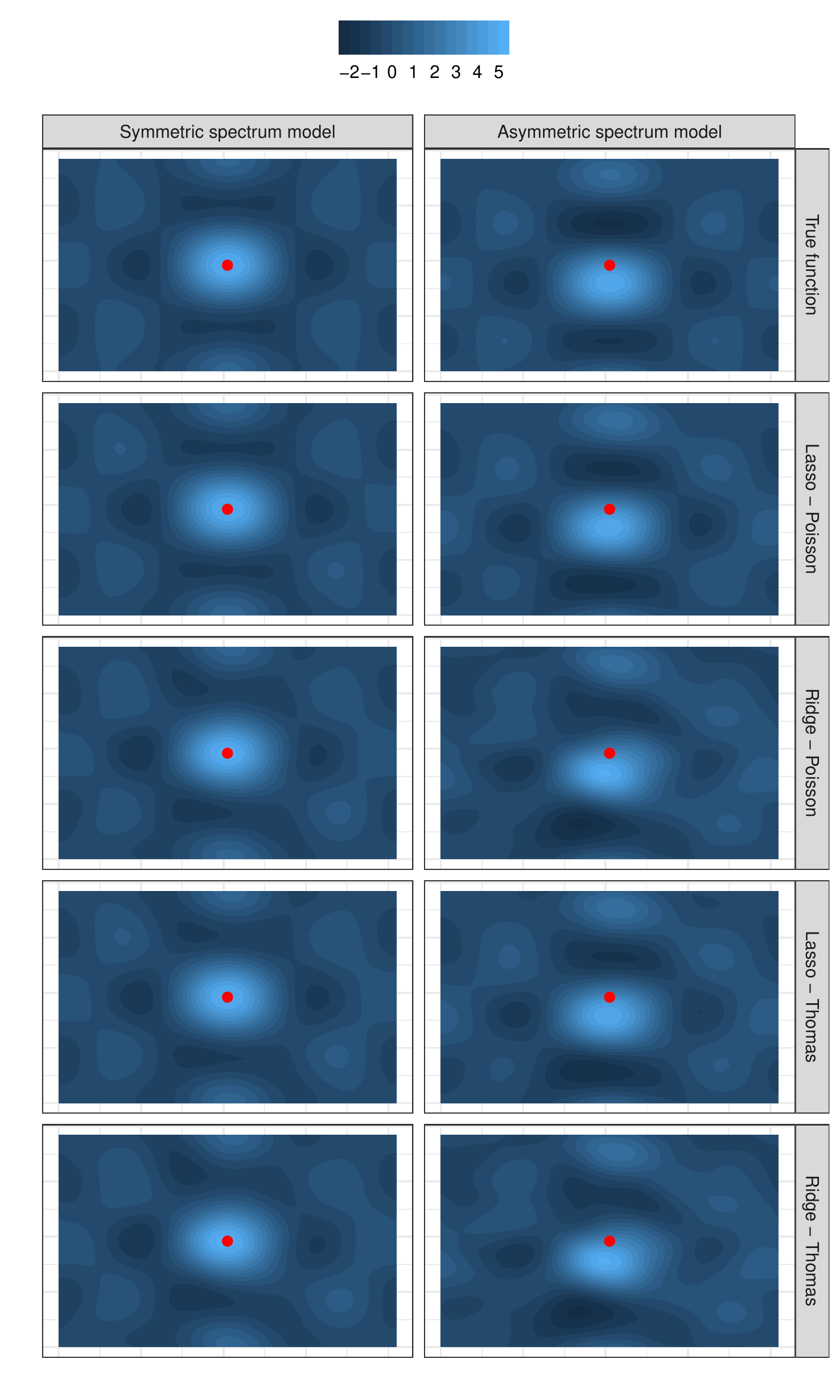}
\caption{True $\beta$ function (first row) and average (rows 2--4) estimates under different models considering $\mathbb{E}\{N(W)\}= 1800$ and $K = 112$. The used $\beta$ function is detailed in the column name whilst the simulated point processes and the used estimation procedure is specified on the label of each row. All plots share the same scale. The black dot is the center of the image (the coordinate (512,393)) and it has been added as a reference.} \label{fig:mean}
\end{figure}

We focus on  planar point processes, set $W = [0,1024]\times[0,786]$ and consider  two point process models and two functions $\beta$ (four scenarios so). As models, we consider an inhomogeneous Poisson point process and an inhomogeneous Thomas process. For both types of processes, we suppose that the intensity function $\rho$ follows the log-convolution model given by~\eqref{eq:intensity} where the function $\beta$ has a compactly supported spectrum. We set $\beta(\boldsymbol{s}) =\beta_{\boldsymbol{\kappa}_0} +  \sum_{i=1}^{K_0} \beta_{\boldsymbol{\kappa}_i}\phi_{\boldsymbol{\kappa}_i}(\boldsymbol{s})$, where $K_0 = 12$ is the true number of frequencies, and $\{\boldsymbol{\kappa}_i \}_{i=0}^{12}\subset \mathbb{Z}^2$ is the partially ordered frequency sequence illustrated on the left hand side of Figure~\ref{fig:spectrum}. We consider the two following spectral models for $\beta$:

\begin{enumerate}
    \item[(a)] $\beta_{\boldsymbol{\kappa}_{i}} = 0.3, i=1,\ldots,12 $. The spectrum is illustrated on the middle of Figure \ref{fig:spectrum}, and it yields a symmetric function $\beta$ which is illustrated on the top left corner of Figure \ref{fig:mean}. 
    \item[(b)] $\beta_{\boldsymbol{\kappa}_{i}} = 0.3 +\mathrm{i} 0.15\kappa_i^{y}, i=1,\ldots,12$ with $\kappa_i^{y}$ being the $y$ coordinate. The real (resp. imaginary) part of the spectrum is illustrated on the middle (resp. right) hand side of Figure \ref{fig:spectrum}, and it yields the asymmetric function $\beta$ depicted on the top right corner of Figure \ref{fig:mean}. 
\end{enumerate}


\begin{figure}[H] 
\centering
\includegraphics[width=12cm]{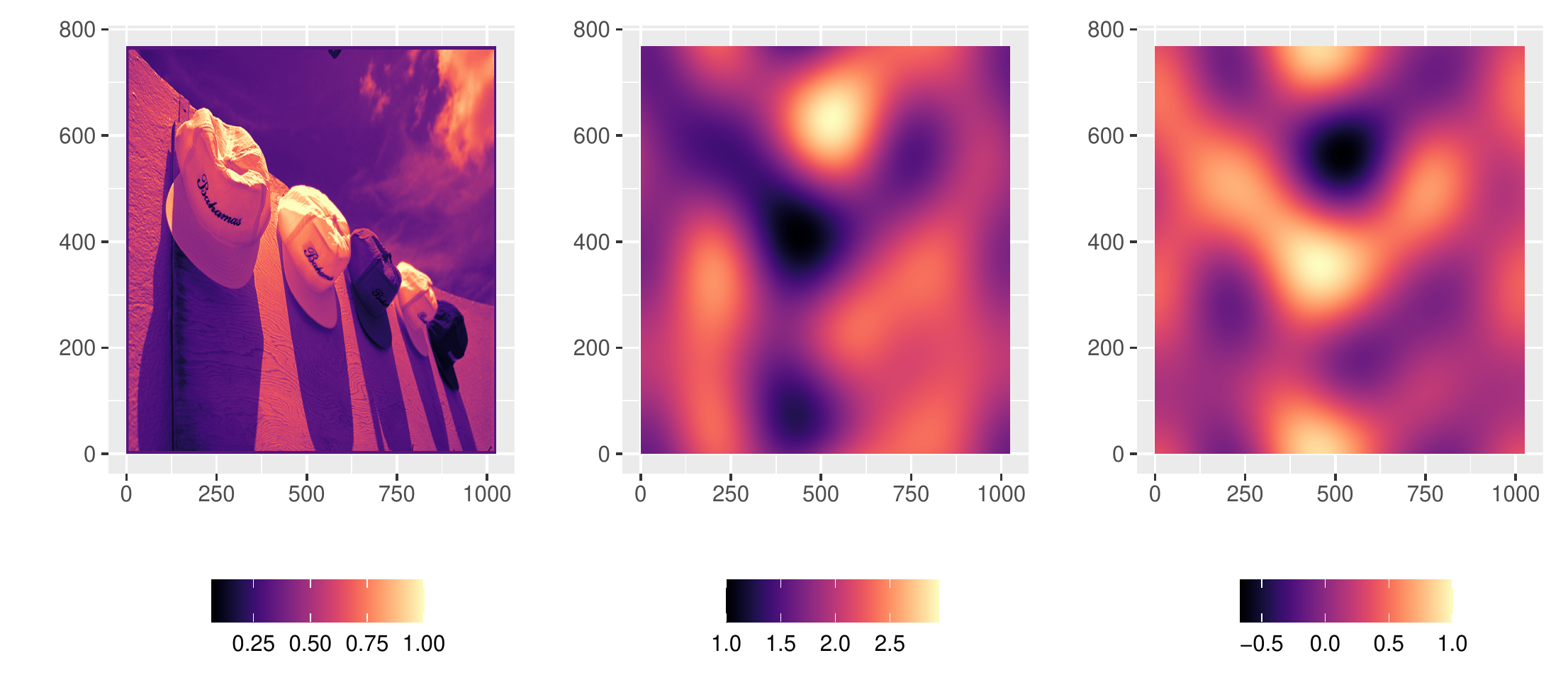}
\caption{Left : True image $Z$. Middle: Resulting convolution between $Z$ and the symmetric function $\beta$ defined in Section \ref{sec:simulation} (a). Right: Resulting convolution between $Z$ and the asymmetric function $\beta$ defined in Section \ref{sec:simulation} (b).} \label{fig:convolution}
\end{figure}

For each scenario, the covariate function $Z$ is the image depicted on the left of Figure \ref{fig:convolution}, and the resulting log-intensity function for the symmetric (a) (resp. asymmetric (b)) function $\beta$ is illustrated on the middle side (resp. right) of Figure \ref{fig:convolution}. Finally, for each scenario, we consider three different values of the parameter $\beta_{\boldsymbol{\kappa}_{0}}=\beta_{\boldsymbol{0}}$, where these values are chosen such that the expected number of points $\mathbb{E}\{N(W) \}$ is $200$, $800$, and $1800$ (to emulate infill asymptotic framework). For the Thomas process, we 
impose 100 clusters in average and set the scale parameter to respect the expected number of points.

For each scenario and each value of $\beta_{\boldsymbol{0}}$, we simulate $1000$ point patterns. For each simulated point pattern, we estimate the spectrum of $\beta$ using the log-linear model given by~\eqref{eq:linear_model3} with the three following methods: 
\begin{itemize}
 \item Poisson likelihood: maximizing the Poisson likelihood function;
\item Ridge: maximizing an adaptive Ridge regularized version of the Poisson likelihood following \cite{choiruddin2018convex};
\item Lasso: Adaptive Lasso estimator defined in the previous section.
\end{itemize}

The vector of covariates, say $\boldsymbol{Z}(\boldsymbol{s})$ in \eqref{eq:linear_model3}, was standardized, i.e. $\|\mathcal{R}[Z_{\boldsymbol{\kappa}_i}\phi_{\boldsymbol{\kappa}_i}(\boldsymbol{s})]\| = 1 = \|\mathcal{I}[Z_{\boldsymbol{\kappa}_i}\phi_{\boldsymbol{\kappa}_i}(\boldsymbol{s})]\|$ for $i=1,\ldots,K$. 

We assume that the true number of frequencies $K_0$, is unknown and we study the performance of the estimates of \eqref{eq:linear_model3} using different values of $K$. Since the partial ordering of the frequencies illustrated in Figure \ref{fig:spectrum} follows a spiral with center at the origin, we obtained the number $K$ of frequencies to be considered from the sequence $\{2(t+1)^2 + 2(t+1)\}_{t=1}^6$ and its middle points $\{(t+1)^2 + (t+2)^2+2t+3\}_{t=1}^5$, for a total of $11$ considered values for $K$ (namely $K=12,18,24,32,40,50,60,72,84,98$ and $112$). The standard Poisson likelihood regression cannot handle a large number of covariates, so to avoid numerical problems, we only use $K = 12, 18$ and $24$ for this estimator. Recall that the total number of estimated parameters is $p = 2K$ plus an intercept. 
Our aim is to illustrate that  even for a large value of $K$, the regularized estimators are quite efficient compared to the oracle which corresponds to the Poisson likelihood estimator with $K=12$.


The implementation has been done using the \texttt{R} statistical software and results in a combination of the \texttt{spatstat} package \citep{baddeley2015spatial} devoted to spatial point pattern analysis and the \texttt{glmnet} package \citep{friedman2010regularization} which provides Lasso and Ridge regularized estimators for GLMs. We also used extensively the discrete Fast Fourier transform \texttt{fft} implemented in the \texttt{base} package.

\begin{figure}[H] 
\centering
\includegraphics[width=\linewidth]{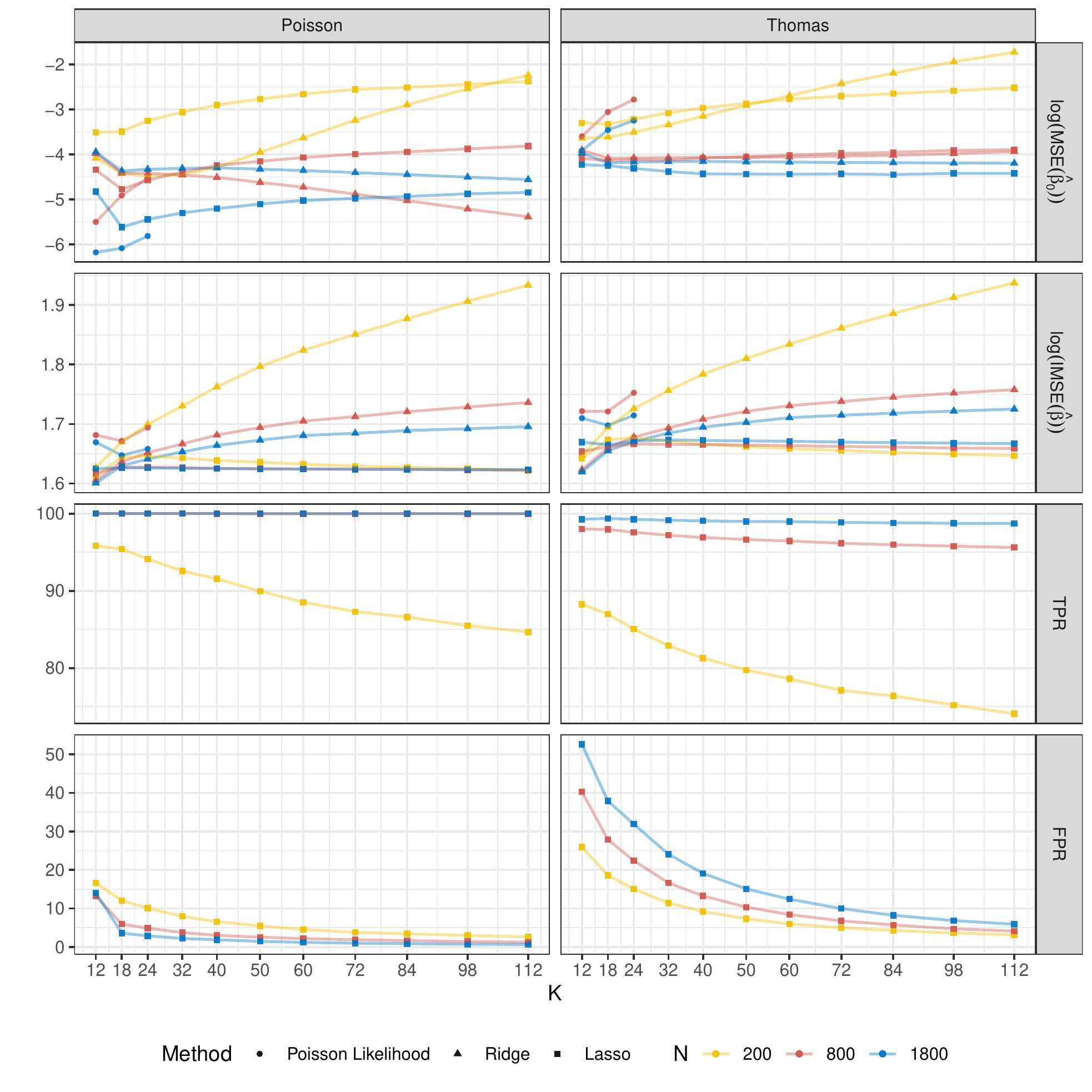}
\caption{Plot of the $\log (\mathrm{MSE})$ (first row), $\log (\mathrm{MISE})$ (second row), $\mathrm{TPR}$ (third row), and $\mathrm{FPR}$ (fourth row) obtained for scenarios where the function $\beta$ is symmetric. The expected total number of points and the used method is detailed in the bottom label, whilst the total number of frequencies $K$ is detailed on the $x$-axis. The simulated point pattern is detailed in each column. The performance of the Poisson likelihood estimate for $N = 200$ is not good so it has been deleted to avoid distortions.} \label{fig:symmetric}
\end{figure}

\begin{figure}[H] 
\centering
\includegraphics[width=\linewidth]{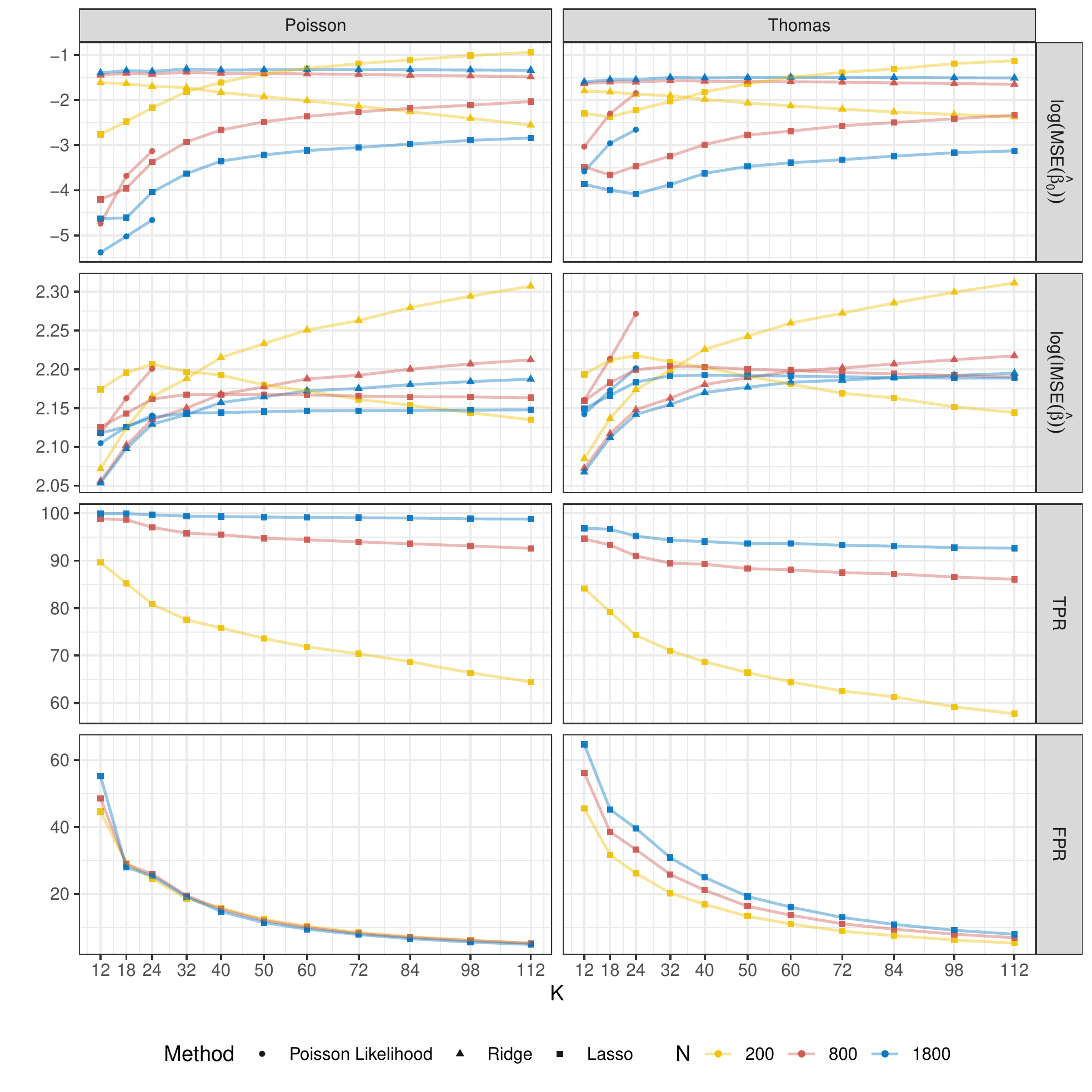}
\caption{Plot of the $\log (\mathrm{MSE})$ (first row), $\log (\mathrm{MISE})$ (second row), $\mathrm{TPR}$ (third row), and $\mathrm{FPR}$ (fourth row) obtained for scenarios where the function $\beta$ is asymmetric. The expected total number of points and the estimation method are detailed in the bottom label, whilst the total number of frequencies $K$ is detailed on the $x$-axis. The simulated point pattern is detailed in each column. The performance of the Poisson likelihood estimates for $N = 200$ is not good so it has been deleted to avoid distortions.} \label{fig:asymmetric}
\end{figure}

In order to evaluate the performance of the different estimation procedures we calculate for each of them the mean squared error (MSE) of $\hat \beta_{\boldsymbol{0}}$, and the mean integrated squared error (MISE) of the function $\hat \beta$ with $\beta_{\boldsymbol{0}}$ set to $0$, that is 

\begin{eqnarray*}
&&\mathrm{MSE} (\hat{\beta}_{\boldsymbol{0}}) = \frac{1}{M}\sum_{m = 1}^{M} \left\{\beta_{\boldsymbol{0}} - \hat{\beta}_{\boldsymbol{0}}^{(m)} \right\}^{2}, \\
&&\mathrm{IMSE}\left\{
\sum_{i = 1}^{K} \hat \beta_{{\boldsymbol{\kappa}_i}}
\phi_{\boldsymbol{\kappa}_i}(\boldsymbol{s}) 
\right\}
 =  \sum_{i = 1}^{K} \mathrm{MSE}(\hat{\beta}_{\boldsymbol{\kappa}_i}),
\end{eqnarray*}

\noindent where $M=1000$ is the number of simulated point patterns and $\hat{\beta}_{\boldsymbol{\kappa}_i}^{(m)}$ is the estimate of $\beta_{\boldsymbol{\kappa}_i}$ for the $m$th point pattern. The first rows of Figures \ref{fig:symmetric} and \ref{fig:asymmetric} contain four graphs, each corresponding to a different simulation scenario. Each curve on a graph is the plot of $\log($MSE) against $K$, for a given method and a given value of the expected number of points. The second rows of these two figures are constructed the same way but for $\log($MISE). We can see that both MSE and MISE decrease with respect to the expected number of points and increase with respect to $K$. However, the increment becomes smaller when the average number of points increases. 

In order to study the selection properties of the adaptive Lasso, we follow \citet{choiruddin2018convex} and report the true positive rate (TPR) given by the number of selected coefficients that are truly different from $0$, hereafter true coefficients, over the total number of true coefficients and the false positive rate (FPR) given by the number of selected coefficients that are truly equal to $0$, hereafter noisy coefficients, over the total number of noisy coefficients. In our context, TPR and FPR play complementary roles: TPR  measures how well the method selects the frequencies associated with true coefficients, while FPR indicates how well the method drops the frequencies corresponding to noisy coefficients. The third and fourth rows of Figures \ref{fig:symmetric} and \ref{fig:asymmetric} give the average TPR and FPR values with respect to the total number of frequencies $K$ for different values of expected number of points. In the two scenarios where the point patterns are realizations of a Poisson point process, we observe that TPR increases and FPR decreases with the expected number of points, meaning that Lasso is able to select the total number of true frequencies and discard the noisy frequencies. When the point patterns are realizations of a Thomas process, the TPR behaves as previously, however, the FPR tends to increase with respect to the expected number of points. 

Overall, what is interesting to point out is that if $K$ is large, which from a practical point of view means we take the maximal number of available frequencies, then the results remain very satisfactory. The MSE and IMSE of respectively $\hat \beta_0$ abd $\hat \beta$ are not too much deteriorated compared to the oracle estimator. Furthermore, the adaptive Lasso procedure is able to recover the true non-zero frequencies efficiently.

Finally, Figure \ref{fig:mean} illustrates the average of the $M=1000$ estimated functions of $\beta$:
$$\hat{\beta}^A(\boldsymbol{s})= M^{-1}\sum_{m=1}^{M} \left\{\sum_{i = 0}^{K} \hat \beta^{(m)}_{{\boldsymbol{\kappa}_i}}\phi_{\boldsymbol{\kappa}_i}(\boldsymbol{s})\right\} $$
obtained with the adaptive Lasso and adaptive Ridge with $K=112$, for our four scenarios and where $\beta_{\boldsymbol{\kappa}_0}$ is set such that the expected number of points is $1800$. 

\section{Data analysis} \label{sec:data}

\begin{figure}[H]
\centering
\includegraphics[scale=.7]{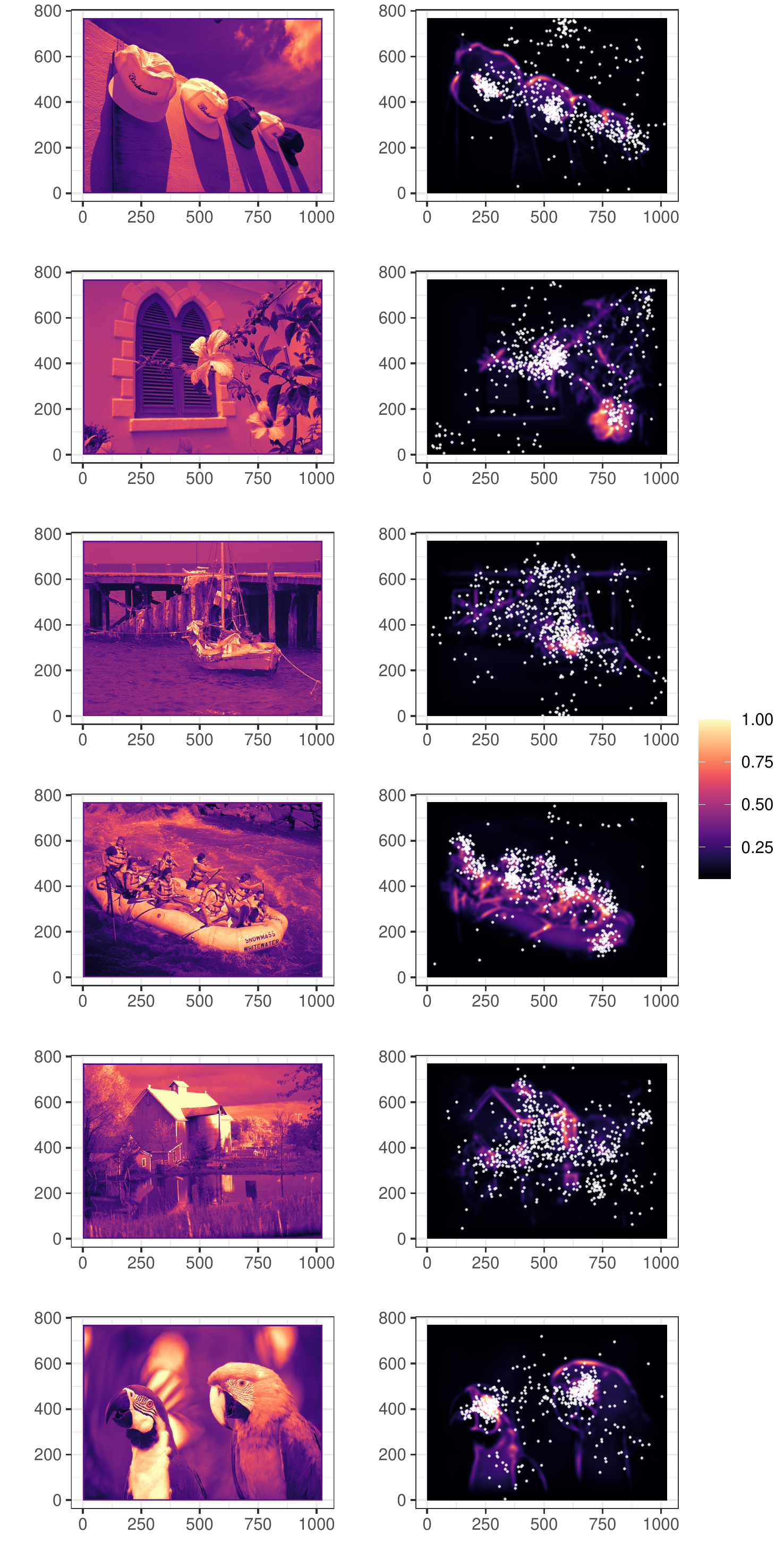}
\caption{\rev{Raw images from our dataset analysis are presented on the first column. The right column depicts saliency maps  obtained from~\citet{ho2010functional} and eye-movement data provided by David Méary and analyzed by~\citep{helo2016eye}. All images share the same scale.}}\label{fig:alldata}
\end{figure}

\rev{In this section, we illustrate our methodology on eye-movement data. Data were kindly provided by David Méary (LPNC, Grenoble) and is a small sample of a larger study conducted on 3- to 12-month old babies and a group of adults to understand visual differences \citep{helo2016eye}. To illustrate this paper, we consider only the group of adults and the six different images that are presented on the first column of Figure~\ref{fig:alldata}. Note that the first image is the same one that has been used for the simulation study (see Figure~\ref{fig:convolution}).
The second column of Figure~\ref{fig:alldata} depicts point patterns, i.e. retina fixations. Points are superimposed on the image of the spatial covariate $Z$ which corresponds to the saliency map, also provided by David Meary, following \citet{ho2010functional}.  We remind that the saliency map is considered to be a deterministic preliminary estimation of the intensity map. One can clearly see from column 2 of Figure~\ref{fig:alldata} that saliency maps fail to capture some regions of images where quite a large number of fixations are recorded. This is also revealed by AUC results of Table~\ref{tab:AUCs} which is more commented hereafter.}

\rev{The objective of this section is to compare several estimates of the intensity function to see if improvements can be made on prediction properties. We consider the following estimators:
\begin{enumerate}
\item Log-linear model: we consider the log-linear regression model~\eqref{eq:loglinear} with $Z$ as covariate and estimate the single parameter $\beta$ by using the Poisson likelihood, then $\hat \rho(\cdot)=\exp(\hat \beta Z(\cdot))$.
\item Semiparametric estimate: we consider the semiparametric model $\rho(s) = f\{Z(s)\}$ and estimate nonparametrically $f$ as detailed in \citet{baddeley2012nonparametric}, then $\hat \rho(\cdot) = \hat f(Z(\cdot))$.
\item Nonparametric estimate: we make no assumption and estimate the intensity using the kernel density estimator \citep[e.g.][]{baddeley2015spatial} where the bandwidth parameter has been selected using the proposal by \citet{cronie2018non}.
\item Log-convolution model: we consider the log-convolution model \eqref{eq:intensity} and estimate the function $\beta$ from its spectrum estimated by adaptive Lasso or adaptive Ridge (both with initial estimate obtained from a Ridge penalization). In both cases, the total number of used frequencies was $112$, sorted as in Figure \ref{fig:spectrum}, and the variables were normalized as in the previous section. Then, $\hat \rho(\cdot)= \exp( \hat \beta_{\mathrm{Lasso}} * Z (\cdot))$ for the adaptive Lasso procedure and $\hat \rho(\cdot)= \exp( \hat \beta_{\mathrm{Ridge}} * Z (\cdot))$ for the adaptive ridge version.
\end{enumerate}}

\rev{For all datasets, the resulting estimates are shown in Figures~\ref{fig:data} and \ref{fig:data2}-\ref{fig:data6} in Appendix respectively. Typically, we can notice that the nonparametric estimate and the ones from the log-convolution model are more or less blurred versions of the image, whereas the parametric and semiparametric estimates are transformed versions of the image. For example for the first image (image of the caps), we can observe that the saliency map in this example does not fully explain locations of fixations in the top of the image which reveals why the parametric and semiparametric estimates also miss these points. 
To evaluate numerically prediction performances of the estimates, we use the area under the ROC curve (AUC)  \citep[Section 6.7.3]{baddeley2015spatial}, also implemented in the \texttt{spatstat R} software. The results are reported in Table \ref{tab:AUCs}. This table first illustrates that all estimates provide a significant improvement compared to the saliency map. It also sheds the light on the log-convolution model. In almost all situations the AUC is higher for these methods (Lasso and Rigde) than for the other ones.}

\begin{figure}[H]
\centering
\includegraphics[scale=.65]{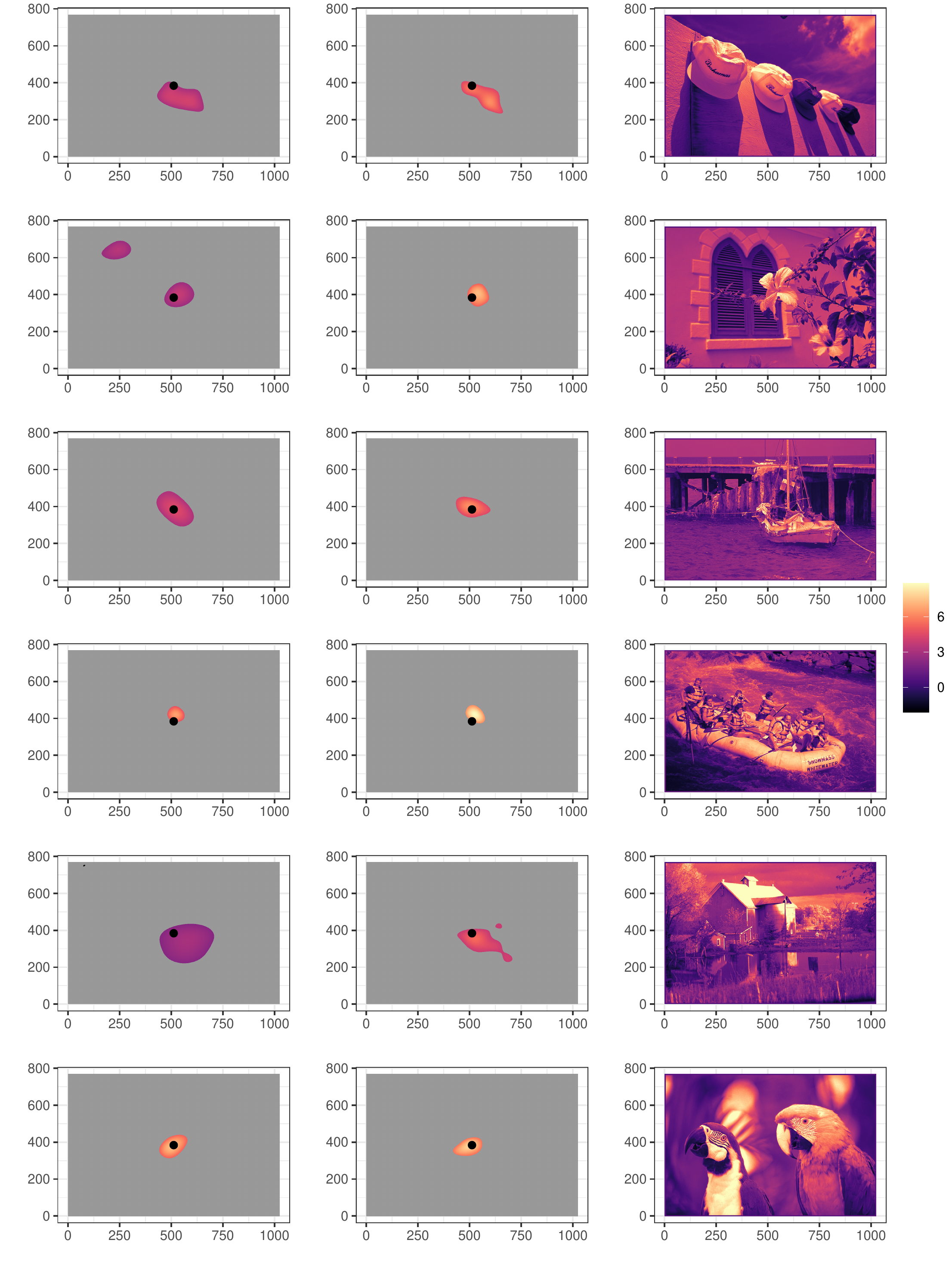}
\caption{\rev{Adaptive Lasso (first column) and Ridge (second column) estimates of the $\beta$ function resulting from the log-convolution model applied to all datasets represented in Figure~\ref{fig:alldata} based on raw images in the last column. First two columns have been truncated to values exceeding 75\% of their absolute maximal value. Note that the scale of the last column corresponds to the one in Figure~\ref{fig:alldata}.}} \label{fig:all_betas}
\end{figure}

\rev{The log-convolution model seems to be a good compromise between the parametric and nonparametric estimates. We have the nonparametric flexibility while keeping some interpretation on the influence of the spatial covariate $Z$. Indeed, we also have access to an estimate of the $\beta$ function. For each dataset the estimated $\beta$ function is shown for the six images in Figure~\ref{fig:all_betas}, and depict how the intensity function on a location $\boldsymbol{s}_{0}$ is proportional to a weighted version of $Z$ at the location $\boldsymbol{s}_{0}$ and its neighbours. To have a better interpretation of these functions, we have truncated the images to values exceeding 75\% of their absolute maximal value. The black dot corresponds to the center of the image and has been added as a reference.}

\rev{Except for the Lasso estimator for Image 2 (Figure~\ref{fig:all_betas}, column 1, row 2), the $\beta(\cdot)$ function reaches its maximum close to the center which is expected as it makes sense that the eye takes the image information in a neighborhood of a retina fixation.
The deviation with respect to the center can be explained by the nature of the image, since the recognizable objects are not always in the center of the observed image. 
Also, we can observe that the dispersion of the kernels is related to the dispersion of point patterns. 
For example, for the rafting image or parrots image (Images 4 and 6), the `interesting' information is very localized on some specific areas. Exploiting more these estimates of the function $\beta$ for different populations or different experimental conditions is definitely an interesting perspective and constitutes the matter of further research.}

\rev{The computations were done in an Intel Dual Xeon E5-2689 computer with 128 GB. Table~\ref{tab:times} provides computing times in seconds to evaluate the log-convolution model using the adaptive ridge and lasso procedures. It makes sense that the lasso estimate is more expensive but the results remain very fast given the large amount of involved computations. The percentage of retained frequencies by the lasso procedure, also shown in Table~\ref{tab:times} sheds light on the capacity of the lasso procedure to discard in average 75\% of frequencies.} 

\newpage

\begin{table}[H]
\rev{\begin{center}
\begin{tabular}{p{.35\linewidth}|llllll}
\hline
& Image 1 & Image 2& Image 3 & Image 4 & Image 5  & Image 6\\ 
& [caps] & [flower] & [boat] & [raft] & [house]  & [parrots]\\ \hline
Number of points (i.e. retina fixations)& 632 & 560 & 582& 631& 608 & 548\\
\hline\hline
Saliency maps: $\hat \rho(\cdot) = Z(\cdot)$ & 0.577& 0.573 & 0.594 & 0.641& 0.538& 0.709 \\       
Parametric method: $\hat \rho(\cdot) = \exp(\hat \beta Z(\cdot))$           & 0.784       & 0.783           & 0.787     & 0.819     & 0.750 & 0.862   \\
Semiparametric  method (Baddeley et al): $\hat \rho(\cdot) =\hat f(Z(\cdot))$& 0.782       & 0.785 & 0.788     & 0.829     & 0.756  & 0.862  \\
Nonparametric estimation     & 0.836       & 0.828          & 0.828     & 0.857     & 0.821    & 0.901\\
Convolution model  (Lasso): $\hat \rho(\cdot)= \exp( \hat \beta_{\mathrm{Lasso}} * Z (\cdot))$    & 0.857       & 0.883          & 0.830     & 0.868     & 0.800   & 0.913 \\
Convolution model  (Ridge):  $\hat\rho(\cdot)= \exp( \hat \beta_{\mathrm{Ridge}} * Z (\cdot))$  & 0.871& 0.862& 0.851& 0.873& 0.824& 0.926 \\
\hline
\end{tabular}
\caption{Results for the areas under ROC curve (AUC) for the different estimates for the set of 6 datasets presented in Figure~\ref{fig:alldata}.} \label{tab:AUCs}
\end{center}}
\end{table}

\begin{table}[H]
\centering
\rev{\begin{tabular}{p{.3\linewidth}|llllll}
\hline
& Image 1 & Image 2& Image 3 & Image 4 & Image 5  & Image 6\\ 
& [caps] & [flower] & [boat] & [raft] & [house]  & [parrots]\\
&&&&&&\\
\hline\hline
Computing time for the ridge estimate (in seconds)& 3.4 & 3.6& 3.4& 4& 3.6& 3.4\\
\hline
Computing time for the lasso estimate (in seconds)& 17.9& 14.2& 16.6& 46.4& 17.6& 22.9\\                   
Percentage of non zero frequencies for the lasso (in \%)& 32& 27& 15& 38& 9& 29\\
\hline
\end{tabular}}
\caption{Computing times to estimate the log-convolution model using the adaptive ridge or lasso method. We also report for the lasso procedure the (rounded) percentage of non zero frequencies retained by the procedure (over a possible total of 112 frequencies).} \label{tab:times}
\end{table}

\begin{figure}[H]
\centering
\includegraphics[scale=.8]{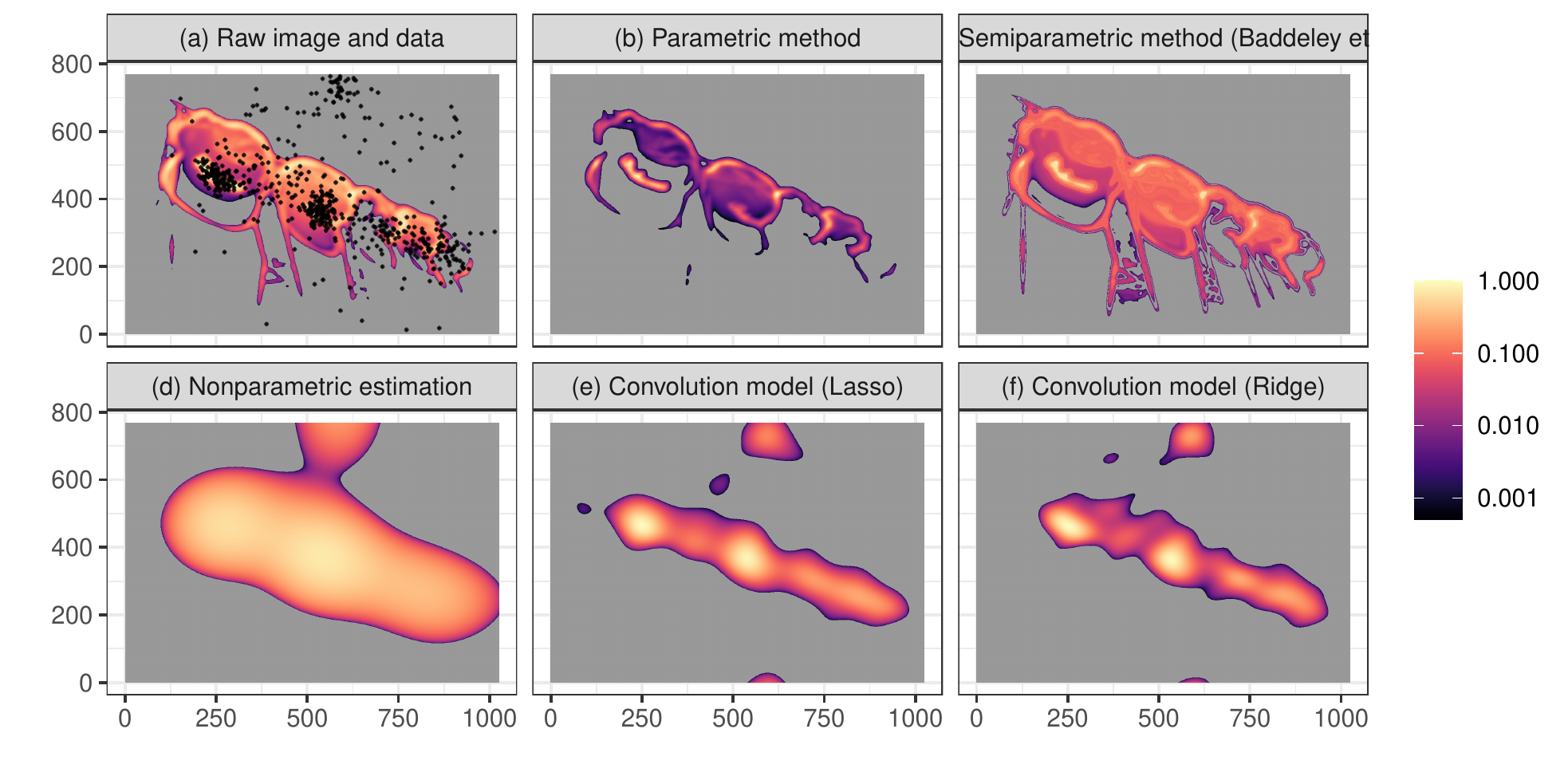}
\caption{Data points and saliency map (covariate $Z$) are represented on the top left. The raw image corresponds to row number 1 of Figure~\ref{fig:alldata}. The other figures represent estimates of the intensity function obtained with parametric (log-linear model), semiparametric and nonparametric methods. Estimates obtained from the log-convolution model (Lasso and Ridge) are also represented. Images are rescaled to $[0,1]$ for a better visualization. All plots share the same scale.} \label{fig:data}
\end{figure}


\section{Discussion}

\rev{This paper has introduced a log-convolution model to parameterize the intensity of a spatial point process. Applied to eye-movement data, for which we often observe retina fixations and a saliency map, this model can be seen as a good compromise between the log-linear model and a nonparametric estimator as we assume that the probability to observe a point is a local average of the saliency map. The model and the proposed procedures to infer it exhibit interesting prediction performances on the one hand and has the flexibility to remain interpretable. To continue this work, it would be really interesting to compare the shapes of estimated functions $\beta(\cdot)$ (for instance in terms of dispersion, anisotropy) for different populations, or for different experimental conditions. We leave this perspective for a future research.}

\rev{Following the will to interpret estimates of the function $\beta(\cdot)$, an interesting and definitely  challenging perspective is to propose a regularization (a fortiori more complex than an adaptive lasso or ridge penalty) of the spectral coefficients $\beta_{\boldsymbol \kappa_i}$ such that not the spectral coefficient but the function $\beta$ itself is ensured to be smooth, decreasing or compactly supported. }

\rev{In the same way, even for such data, we could consider a multivariate version of the log-convolution model to incorporate for instance more characteristics of the raw image. The model could be written as $\log \rho(\bs) = \sum_{i=1}^{I} \beta_{i} * Z_{i}(\bs)$ (for instance for  $I=3$, $Z_i$ could correspond to the RGB levels of the raw image). The methodology proposed in this paper should be easy to extend in order to estimate each function $\beta_i$. However a restriction of the form $\sum_{i = 1}^{I} \int_W \beta_{i}(\bs) d\bs = 1$ would probably be required. The aforementioned restriction would only allow us to recover the shape of each $\beta_i(\cdot)$ but not its zero frequency. Such an extension will probably imply numerical complexities as the number of parameters to estimate will quickly grow up. The estimated $\beta$ function for the proposed model provides more information which can be exploited for practical applications.}

\rev{Finally, we conside in this paper $d$-dimensional point patterns, with a focus on the planar case. We believe that the model and methodology should be straightforwardly extended to circular or spherical point processes. Due to the property of Fourier bases, we even think that the implementation would be easier.}

\section*{Acknowledgements}

The authors would like to thank David Méary for fruitful discussions and for providing us illustrating data. The research of Jean--François Coeurjolly and Marie--Hélène Descary is supported by the Natural Sciences and Engineering Research Council of Canada. Francisco Cuevas--Pacheco has been supported by  ANID/FONDECYT/POSTDOCTORADO/No. 3210453 and the AC3E, UTFSM, under grant FB-0008.


\bibliography{convolSPP.bib}

\appendix


\section{Infill asymptotic results} \label{sec:results}

\subsection{Notation and main result}

\rev{In this section, we present new asymptotic results regarding the estimation of $\boldsymbol \psi$ using an adaptive Lasso regularization of the Poisson likelihood. Remind that $\hat{\boldsymbol{\psi}}=\mathrm{argmin}_{\boldsymbol \psi} Q_n(\boldsymbol \psi)$ where $Q_n$ and $\ell_n$ are given by \eqref{eq:Qn} and \eqref{eq:elln} respectively.
Let us first introduce some additional notation.
Let $\bell_n^{(1)}(\bpsi) \in \mathbb R^p$ be the derivative of $\ell_n(\bpsi)$ with respect to $\bpsi$, it is well-known that the sequence $\{\bell_n^{(1)}(\bpsi)\}_{n\ge 1}$ constitutes a sequence of estimating equations. In that respect, we let $\bpsi_0=\bpsi_{0n}= \mathrm{argmin}_{\bpsi} \mathbb{E}\{\bell_n^{(1)}(\bpsi)\}$ and $\hat \bpsi=\hat \bpsi_n=\mathrm{argmax}_{\bpsi} Q_n(\bpsi)$. Moreover, let $\bS_n(\bpsi)=\mathbb{E}\left\{-\frac{\dd}{\dd \bpsi^\top} \bell_n^{(1)}(\bpsi) \right\} = -\frac{\dd}{\dd \bpsi^\top} \bell_n^{(1)}(\bpsi)$ and $\bSigma_n(\bpsi_0)=
\Var \{\bell_n^{(1)}(\bpsi)\}$ be the $(p,p)$ matrices given by
\begin{align*}
    \bS_n(\bpsi) &= \theta_n \int_W \by(\bs)\by(\bs)^\top \exp(\bpsi^\top \by(\bs)) \dd \bs, \\
\bSigma_n(\bpsi_0)  &= \bS_n(\bpsi_0) +
\theta_n^2 \int_W\int_W \by(\bs)\by(\bt)^\top \{g_n(\bs,\bt)-1 \} \exp(\bpsi_0^\top( \by(\bs)+\by(\bt)) \dd \bs \dd \bt.
\end{align*}
We assume our model is sparse in the sense that $\bpsi_{0}$ can be decomposed as $\bspi_0=(\bpsi_{01}^\top,\bpsi_{02}^\top)^\top$ where $\bpsi_{01}\in \mathbb R^s$, with $s<p$,  and $\bpsi_{02}={\bf 0} \in \mathbb R^{p-s}$ is a zero vector. Following this decomposition we let \linebreak$\by(s)=\{\by_1(s)^\top, \by_2(s)^\top\}^\top$ and $\bpsi=(\bpsi_1^\top,\bpsi_2^\top)^\top$ and more generally for any vector $\bz \in \mathbb R^p$ we use the notation $\bz=(\bz_1^\top, \bz_2^\top)^\top$. For a square $(p,p)$ positive semi-definite matrix $\mathbf M_n$, we let $\nu_{\min}(\mathbf M_n)$ denote its smallest eigenvalue, $\|\mathbf M_n\|$ denote its spectral norm and $\mathbf M_{n,11}$ denote its $(s,s)$ top-left corner. 
Finally, we let 
\begin{equation}
    \label{eq:anbn}
    a_n = \max_{j\le s} \lambda_{n,j} \quad \text{ and } \quad
    b_n = \min_{j>s} \lambda_{n,j}.
\end{equation}
Our main result is based on a set of assumptions that we gather under the assumption [H].\\
\begin{itemize}
    \item[[H]\!\!\!] $\boldsymbol{\Psi}$ is an open bounded convex set of $\mathbb R^p$, $\sup_{s\in W} \|\by(s)\|_\infty<\infty$, \linebreak $\liminf_n {\nu_{\min}\{\theta_n^{-1} \bS_n(\bpsi_0)\}}>0$, $\liminf_n {\nu_{\min}\{\theta_n^{-1} \bSigma_n(\bpsi_0)\}}>0$, $\|\bSigma_n(\bpsi_0)\|=\mathcal O(\theta_n)$ and as $n\to \infty$
    $$ 
    \bSigma_{n,11}^{-1/2}(\bpsi_0) \bell_{n,1}^{(1)}(\bpsi_0) \stackrel{d}{\to} N(0,\mathbf I_s)  
    $$ in distribution.
\end{itemize}  }

\begin{theorem}\label{thm:infill}
We assume the set of assumptions [H] holds and let $\theta_n \to \infty$ as $n\to \infty$. Then we have the two following statements.

(i) Assume $a_n=\mathcal O(\theta_n^{-1/2})$, then $\hat \bpsi-\bpsi_0=\mathcal O_\P(\theta_n^{-1/2})$.

(ii) Assume $a_n \sqrt{\theta_n}\to 0$ and $b_n \sqrt{\theta_n}\to \infty$, then as $n\to \infty$, $\P(\hat \bpsi_2=0)\to 1$ and 
$$
\bSigma_{n,11}^{-1/2}(\bpsi_0) \bS_{n,11}(\bpsi_0) (\hat \bpsi_1 - \bpsi_{01}) \stackrel{d}{\to} N(0,\mathbf I_s)
$$
\indent in distribution.
\end{theorem}

Theorem~\ref{thm:infill} is the natural infill version of  \citet[Theorem~1-2]{choiruddin2018convex} where the volume $|W_n|$ in the increasing domain asymptotics is replaced by $\theta_n$.

\rev{We outline that following \citet{choiruddin2018convex} and \cite{ba2020high}, results could be easily extended to the setting where $p=p_n$ diverges to infinity. In such a setting, we claim that we would require $p_n^4/\theta_n\to 0$, $a_n\sqrt{\theta_n}\to 0$ and $b_n\sqrt{\theta_n/p_n^2}\to \infty$ as $n\to \infty$. Also, similar results could be obtained for more general penalties such as the adaptive ridge, elastic net, including non-convex penalties such as SCAD or MC+ \citep[see again][]{choiruddin2018convex}. This has not been considered in this paper. We only claim here that for the adaptive ridge penalty also considered in the simulation and data analysis, only (i) is valid.}

\subsection{Proof of Theorem~\ref{thm:infill}} \label{sec:proofs}

\begin{proof}
The proof shares several similarities with~\citet[Theorems~1-2]{choiruddin2018convex}, except mainly that $|W_n|$ is replaced by $\theta_n$. We only give a sketch of the proof below. The notation $\kappa$ hereafter stands for a generic positive constant which may vary from line to line.

\rev{(i) Following existing proofs, we have to prove that for any $\varepsilon>0$, there exists $\Omega>0$ such that 
$$
\P \left\{ \sup_{\|\omega\|=\Omega} \Delta_n(\bomega)>0\right\}\le \varepsilon \quad \text{where} \quad 
\Delta_n(\bomega) = Q_n(\bpsi_0+ \bomega /\sqrt{\theta_n}) - Q_n(\bpsi_0)
$$
for $\bomega\in \mathbb R^p$. We decompose $\Delta_n(\bomega)=T_1+T_2$ as
\begin{align*}
    T_1 &= \theta_n^{-1} \left\{ \ell_n(\bpsi_0+ \bomega/\sqrt{\theta_n})-\ell_n(\bpsi_0)\right\},\\
    T_2 &=\sum_{j=1}^p \lambda_{n,j}\left(|\psi_{0j}|-|\psi_{0j}+\omega_j/\sqrt{\theta_n}| \right).
\end{align*}
Using Taylor expansion
\begin{equation*}
    \theta_n T_1 = \theta_n^{-1/2} \bomega^\top \bell_n^{(1)}(\bpsi_0)-\frac12 \theta_n^{-1} \bomega^\top \bS_n(\bpsi_0) \bomega+\frac12 \theta_n^{-1} \bomega^\top  \left\{ \bS_n(\bpsi_0)-\bS_n(\tilde{\bpsi})\right\} \bomega 
\end{equation*}
for some $\tilde{\bpsi}$ on the line segment between $\bpsi_0$ and $\bpsi_0+ \bomega/\sqrt{\theta_n}$. On the one hand by denoting $\check \nu=\liminf \nu_{\min}\{\theta_n^{-1} \bS_n(\bpsi_0)\}$ and by using [H], we have
\[
	\frac12 \theta_n^{-1} \bomega^\top \bS_n(\bpsi_0) \bomega \le -\frac{\check \nu}2 \|\bomega\|^2.
\]
On the other hand using a Taylor expansion and again [H], we can show that
\[
	\frac12 \theta_n^{-1} \bomega^\top  \left\{ \bS_n(\bpsi_0)-\bS_n(\tilde{\bpsi})\right\} \bomega = O(\|\tilde\bpsi - \bpsi_0\|) = O(\theta_n^{-1/2}). 
\]
Hence, for $n$ large enough
\begin{align*}
   \theta_n  T_1 &\leq \theta_n^{-1/2} \bomega^\top \bell_n^{(1)}(\bpsi_0) - \frac{\check \nu}4 \|\bomega\|^2.
\end{align*}
Regarding $T_2$, we have
\[
\theta_n T_2=
\theta_n \sum_{j=1}^s \lambda_{n,j}\left(|\psi_{0j}|-|\psi_{0j}+ \omega_j/\sqrt{\theta_n}| \right) \le \kappa a_n \sqrt{\theta_n} \|\bomega\| \le \kappa \|\bomega\|
\]
by assumption on $a_n$. Hence, for $n$ large enough, we can pick $\Omega$ large enough to ensure that 
$$
\P\left\{ \sup_{\|\omega\|=\Omega} \Delta_n(\bomega)>0\right\} \le 
\P \left\{ \|\bell_n^{(1)}(\bpsi_0)\| \ge \kappa \sqrt{\theta_n} \right\}
$$  
which leads to the result since from [H] it can be derived that $\|\bell_n^{(1)}(\bpsi_0)\|= \mathcal O_\P(\sqrt{\theta_n})$.}

(ii) This statement is proved into two parts. The first one deals with the oracle property while the second one is focused on the asymptotic normality result.

The oracle property is proved if, under the assumption [H] and the conditions of Theorem~\ref{thm:infill}, we prove that with probability tending to $1$, for any {$\boldsymbol \psi_1$} satisfying $\|{\boldsymbol \psi_1 - \boldsymbol \psi_{01}}\|=O_\mathrm{P}(\theta_n^{-1/2})$, and for any constant $K_1 > 0$,
\begin{equation} \label{eq1}
Q_n\Big\{( {\boldsymbol \psi_1}^\top,\mathbf{0}^\top)^\top \Big\}
= \max_{ \| \boldsymbol \psi_2\| \leq K_1 \theta_n^{-1/2}}
Q_n\Big\{({\boldsymbol \psi_1}^\top,{\boldsymbol \psi_2}^\top)^\top \Big\}.
\end{equation}
And to prove~\eqref{eq1}, it is sufficient to show that with probability tending to $1$ as ${n\to \infty}$, for any ${\boldsymbol \psi_1}$ satisfying $\|{\boldsymbol \psi_1 -\boldsymbol \psi_{01}}\|=O_\mathrm{P}(\theta_n^{-1/2})$, for some small $\varepsilon_n=K_1\theta_n^{-1/2}$, and for $j=s+1, \ldots, p$,
\begin{equation}
\label{eq:lem1}
\frac {\partial Q_n(\boldsymbol \psi)}{\partial\psi_j}<0 \quad
\mbox { for } 0<\psi_j<\varepsilon_n \quad \mbox{ and } \quad
\frac {\partial Q_n(\bf \boldsymbol \psi)}{\partial\psi_j}>0 \quad
\mbox { for } -\varepsilon_n<\psi_j<0.
\end{equation}
By Assumption [H], $\| \ell^{(1)}_n(\boldsymbol \psi_0)\|=O_\mathrm{P}(\theta_n^{1/2})$ and there exists $t\in (0,1)$ such that 
\begin{align*}
\frac {\partial \ell_n(\boldsymbol \psi)}{\partial\psi_j}&=\frac {\partial \ell_n{(\boldsymbol \psi_0)}}{\partial\psi_j}+ t {\sum_{l=1}^K \frac {\partial^2 \ell_n\{\boldsymbol \psi_0 + t (\boldsymbol \psi -\boldsymbol\psi_0)\}}{\partial\psi_j \partial\psi_l}}(\psi_l-\psi_{0l}) 
=O_\mathrm{P}(\theta_n^{1/2})+O_\mathrm{P}(\theta_n \theta_n^{-1/2})=O_\mathrm{P}(\theta_n^{1/2}).
\end{align*}
Let $0<\psi_j<\varepsilon_n$ (the other part of~\eqref{eq:lem1} is proved similarly). For $n$ sufficiently large,
\begin{align*}
\mathrm{P} \left ( \frac {\partial Q_n(\boldsymbol \psi)}{\partial\psi_j}<0 \right)&=\mathrm{P} \left ( \frac {\partial \ell_n(\boldsymbol \psi)}{\partial\psi_j} - \theta_n \sign(\psi_j)<0 \right)
\geq \mathrm{P} \left ( \frac {\partial \ell_n(\boldsymbol \psi)}{\partial\psi_j}< \theta_n b_n \right)=1+o(1)
\end{align*}
since $ {\partial \ell_n(\boldsymbol \psi)}/{\partial\psi_j}=O_\mathrm{P}(\theta_n^{1/2})$ and $b_n\theta_n^{1/2} \xrightarrow{} \infty$.

We now focus on the asymptotic normality statement. As shown in (i) and from the oracle property, there is a root-$\theta_n$ consistent local maximizer of $ Q_n\Big\{({\boldsymbol \psi_1}^\top,\mathbf{0}^\top)^\top \Big\}$, which is regarded as a function of  $\boldsymbol {\psi}_1$, and that satisfies
$\frac {\partial Q_n(\boldsymbol {\hat \psi})}{\partial\psi_j}=0 $ for 
$j=1,\ldots,s$
and
$\boldsymbol{\hat \psi}=( \boldsymbol {\hat{\psi}}_1^\top,\mathbf{0}^ \top)^\top.$ We now use a Taylor series expansion. There exists $t\in (0,1)$ and $\boldsymbol{\breve{\psi}}= \boldsymbol{\hat \psi} + t(\boldsymbol\psi_0-\boldsymbol{\hat \psi})$  such that
\begin{align}
0
=&\frac {\partial \ell_n{(\boldsymbol{\hat \psi})}}{\partial\psi_j}- \theta_n \lambda_{n,j} \,\sign(\hat \psi_j) \nonumber\\
=&\frac {\partial \ell_n{(\boldsymbol \psi_0)}}{\partial\psi_j}+{\sum_{l=1}^s \frac {\partial^2 \ell_n{( \boldsymbol \psi_0)}}{\partial\psi_j \partial\psi_l}}({\hat \psi_l}-\psi_{0l})+{\sum_{l=1}^s D_{n,jl}({\hat \beta_l}-\psi_{0l})} - \theta_n\lambda_{n,j} \,\sign(\hat \psi_j)\label{eq:0equal}
\end{align}
where $D_{n,jl}=\frac {\partial^2 \ell_n{(\boldsymbol{\breve{\psi}})}}{\partial\psi_j \partial\psi_l}-\frac {\partial^2 \ell_n{(\boldsymbol \psi_0)}}{\partial\psi_j \partial\psi_l}$. Now, let $\ell^{(1)}_{n,1}(\boldsymbol \psi_{0})$ (resp. $\ell^{(2)}_{n,11}(\boldsymbol \psi_{0})$) be the first $s$ components (resp. $s \times s$ top-left corner) of $\ell^{(1)}_{n}(\boldsymbol \psi_{0})$ (resp. $\ell^{(2)}_{n}(\boldsymbol \psi_{0})= -\mathbf S_n(\boldsymbol \psi_0)$). Let also $\mathbf D_n$ be the $s \times s$ matrix with entries $D_{n,jl}, j,l=1,\ldots,s$, let $\mathbf{s}_n=\{\lambda_{n,1}\sign(\hat\psi_{1}),\ldots,\lambda_{n,s}\sign(\hat\psi_{s}) \}$ and $\mathbf{M}_n= \bSigma_{n,11}^{-1/2}(\boldsymbol \psi_0)$. We premultiply both sides  of~\eqref{eq:0equal} and rewrite them as
\begin{equation}
 \mathbf M_n \left\{ \ell^{(1)}_{n,1}(\boldsymbol \psi_{0})
- \mathbf S_{n,11}(\boldsymbol \psi_0)
 (\hat{\boldsymbol{ \psi}}_1-\boldsymbol \psi_{01})  \right\}
 =  \mathbf M_n \left\{ 
- \boldsymbol D_n (\hat{\boldsymbol{ \psi}}_1-\boldsymbol \psi_{01}) +\theta_n \mathbf{s}_n   \right\}.
 \label{eq:0vec}
\end{equation}
By Assumption [H], conditions of Theorem~\ref{thm:infill} and since $\hat{\boldsymbol\psi}_1$ is a root-($\theta_n$) estimator, we have $\|\mathbf M_n\|=O(\theta_n^{-1/2})$, $\| \boldsymbol D_n\|=O_{\mathrm{P}}(\theta_n^{1/2})$ and $\|\boldsymbol{\hat \psi}_1-\boldsymbol \psi_{01}\|=O_{\mathrm{P}}(\theta_n^{-1/2})$, whereby we deduce first 
$$
\|\mathbf{M}_n \mathbf D_n (\boldsymbol{\hat \psi}_1-\boldsymbol \psi_{01})\|=O_{\mathrm{P}}(\theta_n^{-1/2})=o_{\mathrm{P}}(1) 
\quad \mbox{ and }\quad
\theta_n \,\|\mathbf{M}_n \mathbf s_n\|=O(a_n\, \theta_n^{1/2})=o(1),
$$
and second that $
\mathbf M_n  \ell^{(1)}_{n,1}(\boldsymbol \psi_{0})
- \mathbf M_n \mathbf S_{n,11}(\boldsymbol \psi_0)
 (\boldsymbol{\hat \psi}_1-\boldsymbol \psi_{01})  
=o_{\mathrm{P}}(1)$ which proves the result by Slutsky's Theorem.
\end{proof}


\section{Additional figures}

\begin{figure}[H]
\centering
\includegraphics[scale=.8]{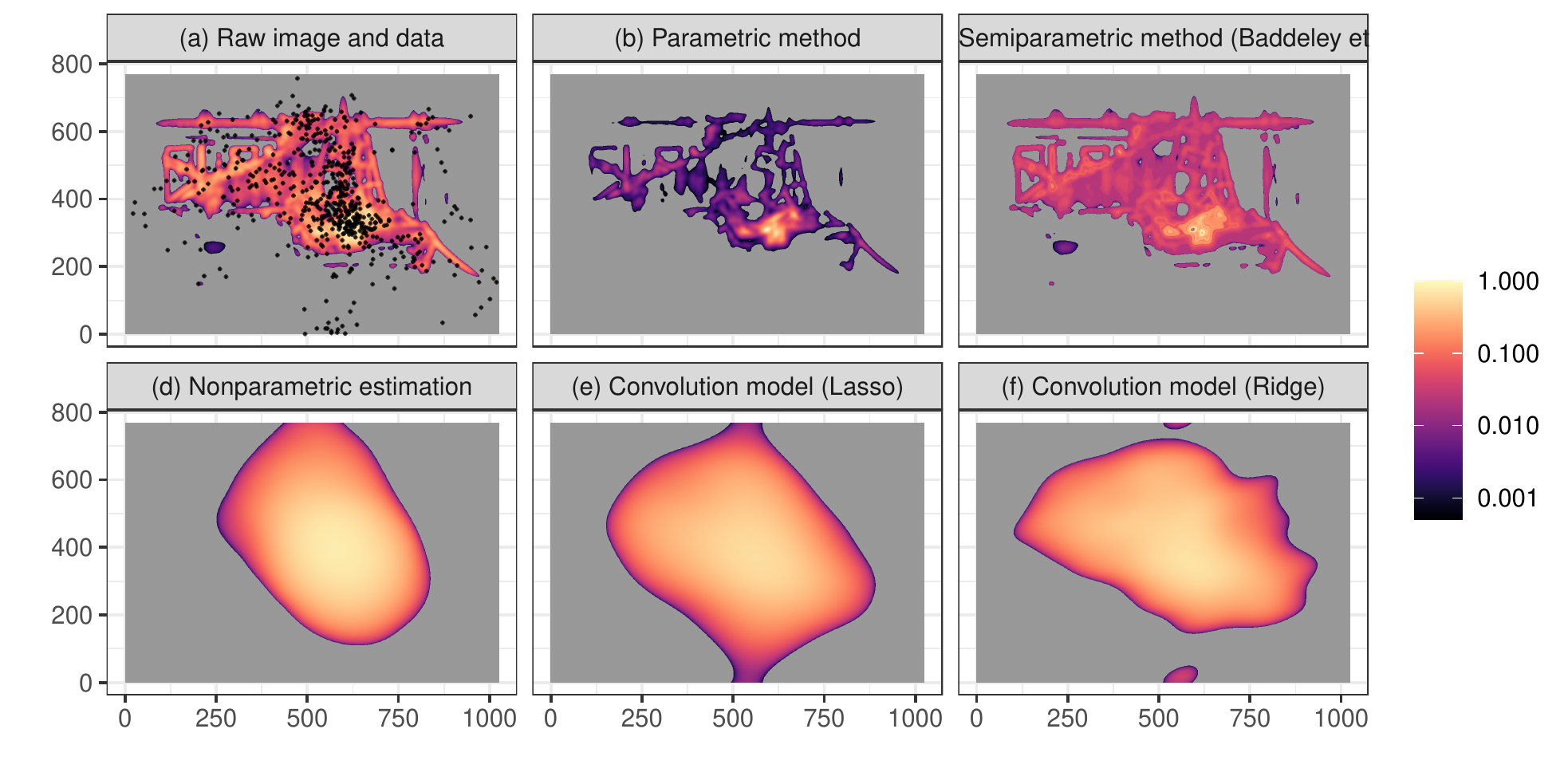}
\caption{Data points and saliency map (covariate $Z$) are represented on the top left. The raw image corresponds to row number 2 of Figure~\ref{fig:alldata}. The other figures represent estimates of the intensity function obtained with parametric (log-linear model), semiparametric and nonparametric methods. Estimates obtained from the log-convolution model (Lasso and Ridge) are also represented. Images are rescaled to $[0,1]$ for a better visualization. All plots share the same scale.} \label{fig:data2}
\end{figure}

\begin{figure}[H]
\centering
\includegraphics[scale=.8]{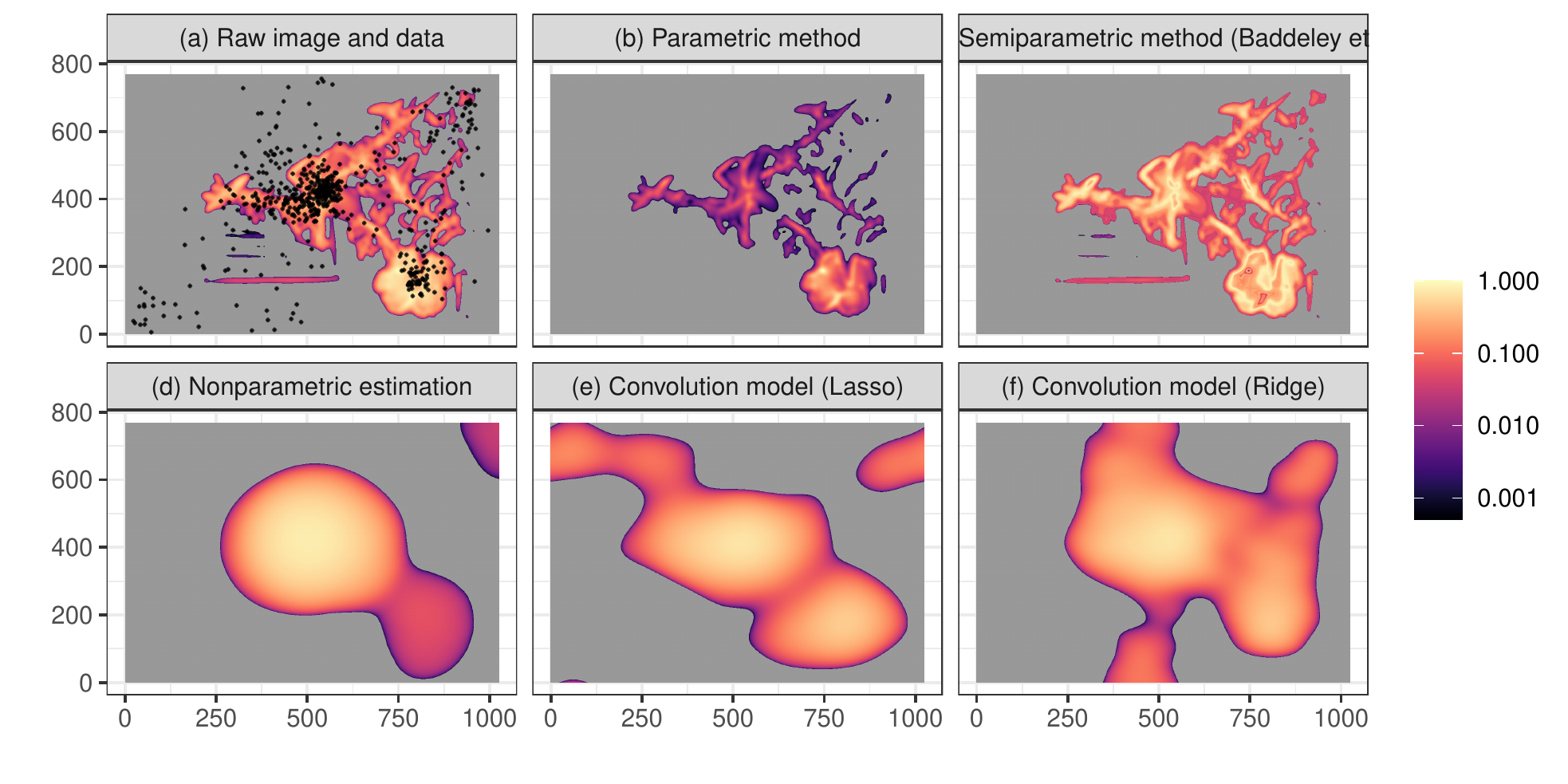}
\caption{Data points and saliency map (covariate $Z$) are represented on the top left. The raw image corresponds to row number 3 of Figure~\ref{fig:alldata}. The other figures represent estimates of the intensity function obtained with parametric (log-linear model), semiparametric and nonparametric methods. Estimates obtained from the log-convolution model (Lasso and Ridge) are also represented. Images are rescaled to $[0,1]$ for a better visualization. All plots share the same scale.} \label{fig:data3}
\end{figure}

\begin{figure}[H]
\centering
\includegraphics[scale=.8]{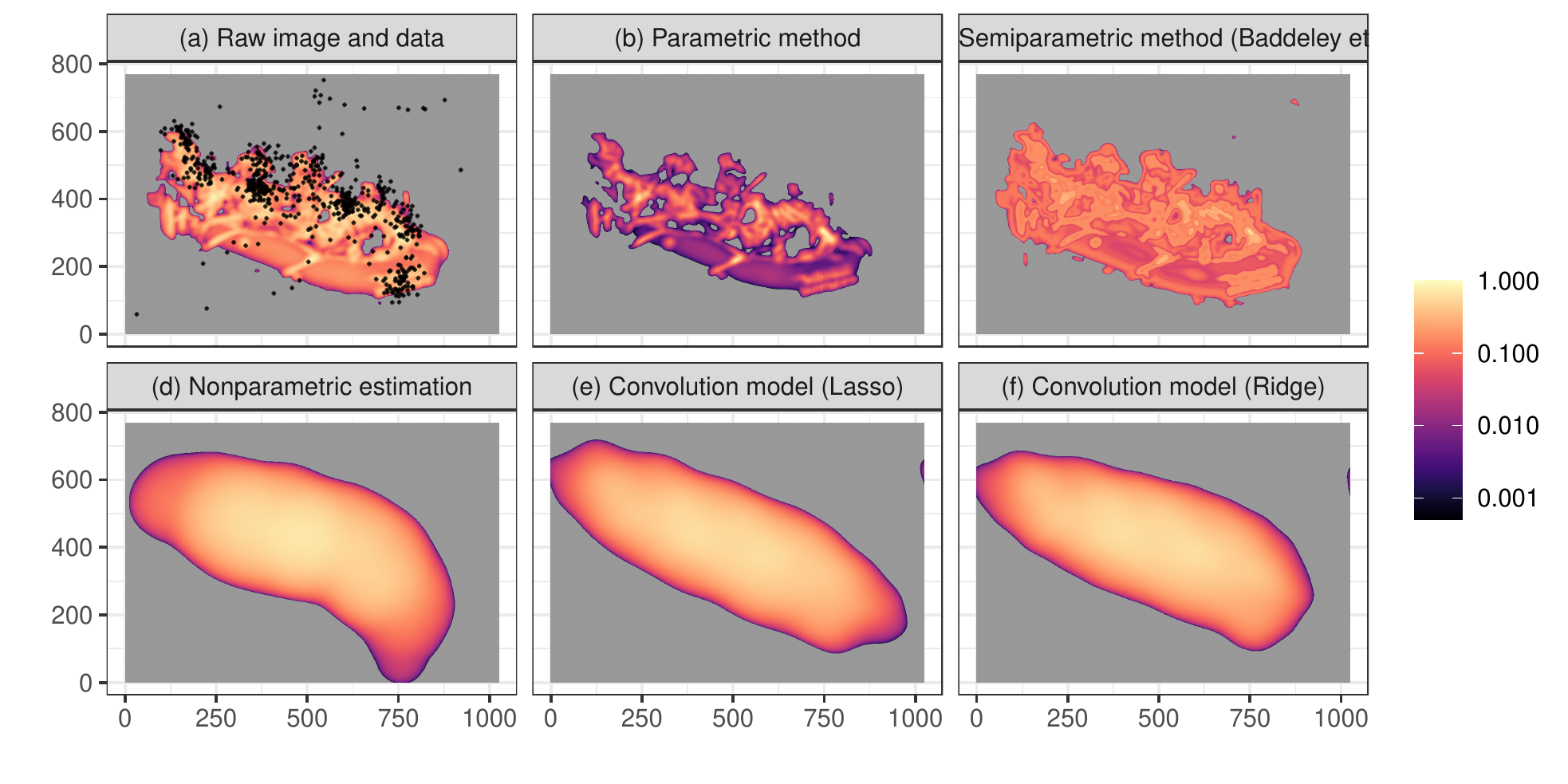}
\caption{Data points and saliency map (covariate $Z$) are represented on the top left. The raw image corresponds to row number 4 of Figure~\ref{fig:alldata}. The other figures represent estimates of the intensity function obtained with parametric (log-linear model), semiparametric and nonparametric methods. Estimates obtained from the log-convolution model (Lasso and Ridge) are also represented. Images are rescaled to $[0,1]$ for a better visualization. All plots share the same scale.} \label{fig:data4}
\end{figure}

\begin{figure}[H]
\centering
\includegraphics[scale=.8]{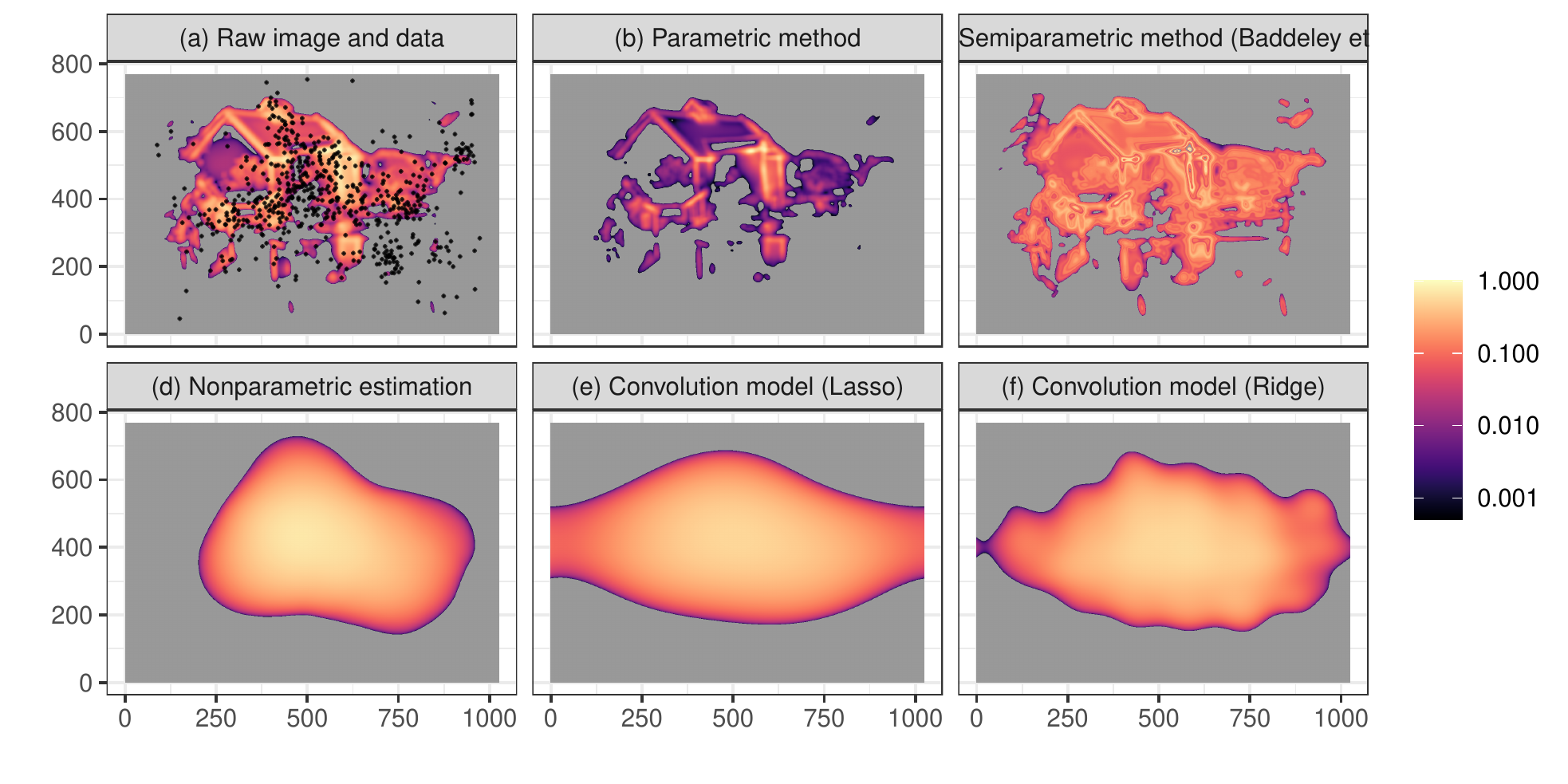}
\caption{Data points and saliency map (covariate $Z$) are represented on the top left. The raw image corresponds to row number 5 of Figure~\ref{fig:alldata}. The other figures represent estimates of the intensity function obtained with parametric (log-linear model), semiparametric and nonparametric methods. Estimates obtained from the log-convolution model (Lasso and Ridge) are also represented. Images are rescaled to $[0,1]$ for a better visualization. All plots share the same scale.} \label{fig:data5}
\end{figure}

\begin{figure}[H]
\centering
\includegraphics[scale=.8]{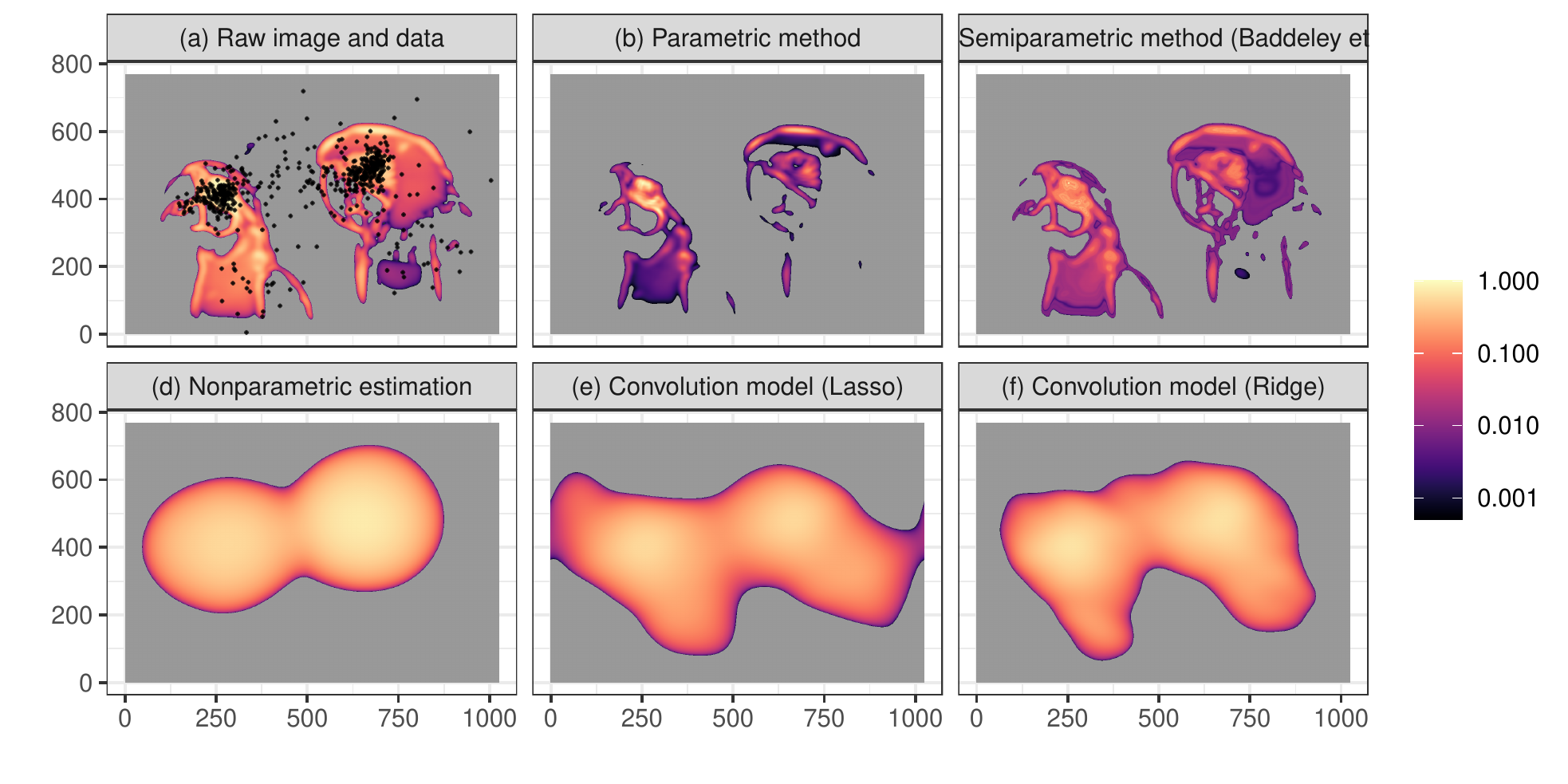}
\caption{Data points and saliency map (covariate $Z$) are represented on the top left. The raw image corresponds to row number 6 of Figure~\ref{fig:alldata}. The other figures represent estimates of the intensity function obtained with parametric (log-linear model), semiparametric and nonparametric methods. Estimates obtained from the log-convolution model (Lasso and Ridge) are also represented. Images are rescaled to $[0,1]$ for a better visualization. All plots share the same scale.} \label{fig:data6}
\end{figure}

\end{document}